\documentclass[reprint,aip,amsmath,amssymb,floatfix,jcp]{revtex4-1}

\usepackage{graphicx}
\usepackage{dcolumn}
\usepackage{bm}

\usepackage[utf8]{inputenc}
\usepackage[T1]{fontenc}
\usepackage{mathptmx}
\usepackage[mathscr]{euscript}
\usepackage{xcolor}

\begin{document}

\title{Computing equilibrium free energies through a nonequilibrium quench}

\author{Kangxin Liu}
\affiliation{Department of Chemistry, New York University}
\affiliation{Simons Center for Computational Physical Chemistry, New York University}

\author{Grant M. Rotskoff}
\affiliation{Department of Chemistry, Stanford University}

\author{Eric Vanden-Eijnden}
\affiliation{Courant Institute of Mathematical Sciences, New York University}
\email{eve2@nyu.edu}

\author{Glen M. Hocky}
\email{hockyg@nyu.edu}
\homepage{http://hockygroup.hosting.nyu.edu/}
\affiliation{Department of Chemistry, New York University}
\affiliation{Simons Center for Computational Physical Chemistry, New York University}

\date{\today}

\begin{abstract}
 Many methods to accelerate sampling of molecular configurations are based on the idea that temperature can be used to accelerate rare transitions. 
 These methods typically compute equilibrium properties at a target temperature using reweighting or through Monte Carlo exchanges between replicas at higher temperatures.
 A recent paper\cite{rotskoff2019dynamical} demonstrated that accurate equilibrium densities of states can also be computed through a nonequilibrium ``quench'' process, where sampling is performed at a higher temperature to encourage rapid mixing and then quenched to lower energy states with dissipative dynamics.
 Here we provide an implementation of the quench dynamics in LAMMPS and evaluate a new formulation of nonequilibrium estimators for the computation of partition functions or free energy surfaces (FESs) of molecular systems.
 We show that the method is exact for a minimal model of $N$-independent harmonic springs, and use these analytical results to develop heuristics for the amount of quenching required to obtain accurate sampling.
We then test the quench approach on alanine dipeptide,  where we show that it gives an FES that is accurate near the most stable configurations using the quench approach but disagrees with a reference umbrella sampling calculation in high FE regions.
 We then show that combining quenching with umbrella sampling allows the efficient calculation of the free energy in all regions.
 Moreover, by using this combined scheme, we obtain the FES across a range of temperatures at no additional cost, making it much more efficient than standard umbrella sampling if this information is required. 
 Finally, we discuss how this approach can be extended to solute tempering and demonstrate that it is highly accurate for the case of solvated alanine dipeptide without any additional modifications.
\end{abstract}

\maketitle

\section{INTRODUCTION}
A major challenge in molecular dynamics (MD) simulations is poor sampling of conformational landscapes because free energy barriers that are large relative to $k_{\rm B} T$ are traversed at rates much lower than the duration of a typical simulation~\cite{frenkel2001understanding,tuckerman2010statistical}.
A wide variety of approaches relying on sampling of equilibrium distributions have been proposed \cite{henin2022enhanced} to circumvent this problem, which can be generally classified into (a) those which seek to lower free energy barriers by adding a bias or changing the potential that is being sampled\cite{torrie1977nonphysical,barducci2008well}, or (b) those that use higher temperatures of all or some degrees of freedom to accelerate transitions\cite{sugita1999replica,liu2005replica}.
Other approaches harness nonequilibrium fluctuation theorems to estimate equilibrium free energies by averaging over many realizations of a nonequilibrium transformation \cite{jarzynski1997equilibrium,vaikuntanathan2008escorted,nilmeier2011nonequilibrium}.
However, these non-equilibrium approaches have not been widely adopted for chemical problems because they are difficult to converge, due to the large variance in work performed and the largest contributions to the equilibrium average being dominated by rare fluctuations.

Ref.~\onlinecite{rotskoff2019dynamical} proposes a class of estimators based on an exact reweighting of the samples gathered during a nonequilibrium process that follows a dissipative dynamical scheme, starting with configurations that are well sampled from an equilibrium density, e.g. the Boltzmann distribution (see also Ref.~\citenum{martiniani2017complexity}, Ch. 5 and Refs.~\citenum{thin_neo_2021} and~\citenum{cao_learning_2022}). 
Conceptually, facile sampling at high temperature allows mixing between free energy basins and then the process of `quenching' allows one to map out the lower energy portion of the free energy basins. 
This method possesses several advantages; it is unbiased and only requires the knowledge of the starting probability density $\rho_0$ up to a constant, unlike the annealed importance sampling\cite{neal2001annealed} (AIS) method, which requires computing a posterior ratio of sample means.
Furthermore, it was proven in Ref~\citenum{rotskoff2019dynamical}  that the estimator has lower variance than a direct estimator and simulations can be run in parallel, which makes the methodology naturally suited to the architecture of high-performance computing clusters. 
The aim of this paper is to investigate how to use this method to compute partition functions and FESs for molecular systems.

To this end, we derive a formulation of these nonequilibrium quench estimators for molecular systems and critically assess the efficacy of this approach.
We demonstrate that the method is exact for a harmonic system where analytical results are available, and, in doing so, we obtain heuristic rules for the required amount of sampling. 
We then demonstrate on the simple test system of alanine dipeptide that, because the method emphasizes low free energy regions, it is not a competitive approach for computing full free-energy surfaces. 
Nevertheless, we find that a combination of quenching with umbrella sampling (US) provides a highly efficient way to compute a full FES for this system, simultaneously giving FESs at many different temperatures. 
Finally, we expand upon the quench method and show that it can be used as a nonequilibrium solute-tempering approach with highly accurate results for a solvated peptide. 
The quenching dynamics are implemented in LAMMPS \cite{plimpton1995fast,thompson2022lammps}, and analysis methods provided as open source Python scripts, meaning that our method can be easily deployed on other problems.

\section{THEORY AND METHODS}
\subsection{An equilibrium estimator from nonequilibrium trajectories}

Let us denote the microstate of the system in phase space by the vector ${\bf x}=({\bf q},{\bf p})\in\mathbb{R}^{2d}$ and ${\bf q},{\bf p}\in\mathbb{R}^{d}$, where $d\equiv 3n$ is the number of degrees of freedom (DOF) in the $n$-atom system.
An average observable property $\phi$ of the system can by computed as an integral over all possible configurations, weighted by the equilibrium probability density\cite{frenkel2001understanding,tuckerman2010statistical},
\begin{equation}
    \langle \phi \rangle = \frac{\int d \mathbf{x}\ \phi(\mathbf{x}) \rho(\mathbf{x})}{\int d \mathbf{x}\ \rho(\mathbf{x})}
    \label{eq:eqaverage}
\end{equation}
A typical example of $\rho(\mathbf{x})$ would be the Boltzmann distribution, 
\begin{equation}
    \rho_\beta(\mathbf{x})=\frac{ e^{-\beta \mathscr{H}(\mathbf{x})} }{\int d\mathbf{x}\ e^{-\beta \mathscr{H}(\mathbf{x})}}\equiv \frac{ e^{-\beta \mathscr{H}(\mathbf{x})} }{Q(\beta)},
    \label{eq:boltzmann}
\end{equation}
where $\beta=\frac{1}{k_B T}$, $k_B$ is Boltzmann's constant, $T$ is the temperature, and $\mathscr{H}$ is the Hamiltonian (total energy function) of the system \cite{frenkel2001understanding,tuckerman2010statistical}.
Here, $Q(\beta)$ is the canonical partition function.

To compute equilibrium averages using molecular dynamics, we typically replace the expectation~\eqref{eq:eqaverage} by a time average along a trajectory ${\bf X}^e(t)$ along which configurations appear in proportion to $\rho_\beta(\mathbf{x})$. 
This can be done using Markov chain Monte Carlo or thermostated molecular dynamics \cite{frenkel2001understanding,tuckerman2010statistical}. 
In this case,
\begin{equation}
    \langle \phi \rangle \approx \frac{1}{\tau} \int_0^\tau dt\phi(\mathbf{X}^e(t)) 
    \label{eq:timeaverage}
\end{equation}
with equality in the limit as $\tau \to\infty$.

As an alternative to running a very long trajectory, if we already had configurations well sampled from $\rho(\mathbf{x})$, then we could also compute this same average by  an ``initiate-and-propagate" procedure, where we draw starting points ${\bf X}_i^e(0)$ from $\rho({\bf x})$ and propagate our equations of motion to get $N$ trajectories of length $\tau_\mathrm{short}$, $\{\mathbf{X}^e_i(t)$\}, computing observables as an average over the trajectory and over initial configurations,
\begin{equation}
    \langle \phi \rangle \approx \frac{1}{N} \sum_{i=1}^N \frac{1}{\tau_\mathrm{short}} \int_0^{\tau_\mathrm{short}} dt \phi(\mathbf{X}^e_i(t)) 
    \label{eq:initiateaverage}
\end{equation}
with equality in the limit as $N\to\infty$ for any $\tau_\mathrm{short}>0$.
The advantage of such a procedure would be that each trajectory can be simulated independently, making the algorithm trivially parallelizable.

Now suppose we wanted to do such a procedure, but the dynamics do not sample the stationary distribution $\rho$, using e.g. the differential equation
\begin{equation}
    \dot{\bf X}(t)=\boldsymbol{b}({\bf X}(t)).
\label{eq:EOM}
\end{equation}
where $\boldsymbol{b}({\bf x})$ is a vector-field to be specified that does not preserves $\rho({\bf x})$, i.e. $\boldsymbol{\nabla} \cdot \left(\boldsymbol{b}({\bf x})\rho({\bf x})\right) \not=0$ where $\boldsymbol{\nabla}$ corresponds to the phase space gradient $\{\frac{\partial}{\partial q_1},\frac{\partial}{\partial q_2},...,\frac{\partial}{\partial q_d},\frac{\partial}{\partial p_1},\frac{\partial}{\partial p_2},...,\frac{\partial}{\partial p_d} \}$.
The change in phase space volume associated with a nonequilibrium dynamical process is quantified by a Jacobian factor, 
\begin{equation}
    J(t)=\exp{\left(\int_0^t{\boldsymbol{\nabla} \cdot \boldsymbol{b}({\bf X}(s)) \,ds}\right)},
\label{eq:J}
\end{equation}
as derived in Appendix~\ref{sec:phasevolume}.

 Ref.~\onlinecite{rotskoff2019dynamical} makes use of the fact that Eq.~\ref{eq:initiateaverage} can be extended to this more general case of motion generated by $\boldsymbol{b}$ by introduction of the density scaled by this Jacobian factor, as long as points can be sampled from the initial density $\rho({\bf x})$.
In this case, estimates are computed for a subset of all phase space by propagating the non-equilibrium trajectories until they reach the boundaries of that subset of phase space, which in practice was done by terminating trajectories at fixed maximum and minimum energy values $E_\mathrm{max}$ and $E_\mathrm{min}$. 
The resulting estimator over $N$ trajectories is given by,
 
\begin{equation}
    \left \langle \phi \right \rangle \approx \lim_{N \to \infty}{\frac{1}{N}\sum_{i=1}^{N}{\frac{\int_{\tau_i^-(E_\mathrm{max})}^{\tau_i^+(E_\mathrm{min})}{dt \phi({\bf X}_i(t)) \rho({\bf X}_i(t)) J(t) }}{\int_{\tau_i^-(E_\mathrm{max})}^{\tau_i^+(E_\mathrm{min})}{dt \rho({\bf X}_i(t)) J(t)}}}}.
\label{eq:general_expression}
\end{equation}
where $\tau_i^+(E_\mathrm{min})$ and $\tau_i^-(E_\mathrm{max})$ are the times that trajectory $i$ reached the fixed energy boundaries when propagating the non-equilibrium dynamics forwards and backwards in time (we emphasize that here the integration times vary for each starting point).

As in Ref.~\onlinecite{rotskoff2019dynamical}, we will use the equations of motion corresponding to zero temperature Langevin dynamics, which we term ``quench'', 
\begin{equation}
    \left \{ \begin{aligned}
    \dot{{\bf Q}} & = {\bf M}^{-1}{\bf P} \\
    \dot{{\bf P}} & = -\boldsymbol{\nabla}{U({\bf Q})}-\gamma {\bf P}
    \end{aligned} \right.
\label{eq:quench_EOM}
\end{equation}
for which the Jacobian is
\begin{equation}
    J(t)=\exp{(-d\gamma t)}.
\label{eq:quench_J}
\end{equation}

This method is easy to implement for molecular systems by adapting the BAOAB scheme \cite{leimkuhler2013robust} (see Appendix ~\ref{sec:langevin}).

Backwards-in-time trajectories from initial points are generated by following the same dynamical scheme using a negative $\gamma$, after reversing the initial momenta. 
This is derived by applying the equations of motion to the time-reversed phase space coordinates ${\bf{Q}}^R(t)={\bf{Q}}(-t)$ and ${\bf{P}}^R(t)=-{\bf{P}}(-t)$.

Because, Eq.~(\ref{eq:quench_EOM}) is dissipative, we can use it to propagate trajectories from high energy to low energy, or from low energy to high energy using a negative $\gamma$.
With this scheme, we can compute Boltzmann averages at the starting inverse temperature $\beta_0$ by first sampling from $\rho_{\beta_0}({\bf x})\propto\exp{(-\beta_0\mathscr{H}({\bf x}))}$ and propagating $N$ trajectories with our quench algorithm forwards and backwards in time, using the formula
\begin{equation}
    \left \langle \phi  \right \rangle_0 \approx 
    \frac{1}{N} \sum_{i=1}^N \frac{ \int_{\tau_i^{-}}^{\tau_i^{+}}dt \phi (\mathbf{X}_i(t)) e^{-\beta_0 \mathscr{H}(\mathbf{X}_i(t) )-d\gamma t} }
    {\int_{\tau_i^{-}}^{\tau_i^{+}} dt e^{-\beta_0 \mathscr{H}(\mathbf{X}_i(t))-d\gamma t}  }
    \label{eq:quench_expression}
\end{equation}
Here we no longer indicate the dependence of $\tau_i^{\pm}$ on energy for brevity.
An extension of Eq.~\ref{eq:quench_expression} to calculate averages at other temperatures above and below $\beta_0$ will be discussed in the next section.

We note that Eq.~\ref{eq:quench_expression} is a biased estimator since it computes expectations over $\rho_{\beta_0}$ conditional on $\mathscr{H}({\bf x})\in [E_\mathrm{min},E_\mathrm{max}]$. To make this bias negligible, we can adjust the values of $E_\mathrm{min}$ and $E_\mathrm{max}$. To this end, notice that  during a (forward) quench, the value of total energy $E_\mathrm{tot}=\mathscr{H}({\bf X}(t))$ will decrease while $d \gamma t$ will increase, resulting in a time where the arguments of the exponentials are maximized that depends on  $\beta_0$. In order to get a converged average, this time must be contained within the range $(\tau^-,\tau^+)$, and so the energy levels, in particular $E_\mathrm{min}$, must be chosen such this is the case.
Our method for doing so is discussed in Sec.~\ref{sec:results}.

\subsection{Calculations of free energies and partition functions}
As mentioned earlier, the principal challenge of computing quantities from MD simulations is that high (free) energy barriers at a temperature of interest prevent proper sampling of all relevant configurations with proper weights. 
Since it can be easier to sample at high temperature, it is tempting to sample at high temperature, and directly reweight samples to lower temperature; for example, to estimate $Q(\beta)$ we can write

\begin{equation}
\begin{aligned}
    Q(\beta) &= \int d \mathbf{x} e^{-\beta \mathscr{H}({\bf x})} = \int d \mathbf{x} e^{-\beta \mathscr{H}({\bf x})} \left(\frac{e^{-\beta_0 \mathscr{H}({\bf x})}}{e^{-\beta_0 \mathscr{H}({\bf x})}}\right) \\
             &= \int d \mathbf{x} e^{(\beta_0-\beta) \mathscr{H}({\bf x})} e^{-\beta_0 \mathscr{H}({\bf x})}.
\end{aligned}
\end{equation}
Hence, we can reweight samples from $\beta_0$ to evaluate the relative value of $Q(\beta)$,
\begin{equation}
    \frac{Q(\beta)}{Q(\beta_0)} = \langle e^{(\beta_0-\beta)\mathscr{H})} \rangle_0,
     \label{eq:Q_ratio}
\end{equation}
which is the central idea of free energy perturbation~\cite{frenkel2001understanding}.
The challenge is that samples from $\rho_{\beta_0}({\bf X})$ do not have good overlap with $\rho_{\beta}({\bf X})$ unless $\beta\approx\beta_0$, so this estimate could have high variance in practice, which can be mitigated using simulated annealing or simulated tempering~\cite{kirkpatrick1983optimization,aluffi1985global,marinari1992simulated}.

The same quantity in Eq.~\ref{eq:Q_ratio} can be computed with the quench estimator (Eq.~\ref{eq:quench_expression}), 
\begin{equation}
    \frac{Q(\beta)}{Q(\beta_0)} \approx {\frac{1}{N}\sum_{i=1}^{N}{\frac{\int_{\tau_i^{-}}^{\tau_i^{+}}{dt e^{-\beta \mathscr{H}({\bf X}_i(t))-d\gamma t} }}
    {\int_{\tau_i^{-}}^{\tau_i^{+}}{dt e^{-\beta_0 \mathscr{H}({\bf X}_i(t))-d\gamma t}}}}}.
\label{eq:Q_ratio_quench}
\end{equation}
By quenching forwards in time, low energy samples which are more relevant at a lower temperature are generated, which should result in a much more robust calculation than reweighting from samples generated only from $\beta_0$ at equilibrium.
Using a similar manipulation, we can compute the average of any observable $\phi(\mathbf{X})$ at  $\beta$ using samples generated from $\beta_0$,
\begin{equation}
\begin{aligned}
    \langle \phi \rangle &= {\langle{\phi e^{(\beta_0-\beta)\mathscr{H}}}\rangle_0} / {\langle{e^{(\beta_0-\beta)\mathscr{H}}}\rangle_0}\\
    &\approx \frac{1}{N} \sum_{i=1}^N \frac{ \int_{\tau_i^{-}}^{\tau_i^{+}}dt  \phi(X_i(t)) e^{-\beta \mathscr{H}(\mathbf{X}_i(t) )-d\gamma t} }
    {\int_{\tau_i^{-}}^{\tau_i^{+}} dt e^{-\beta_0 \mathscr{H}(\mathbf{X}_i(t))-d\gamma t}  } \left(\frac{Q(\beta)}{Q(\beta_0)}\right)^{-1},\\
    \label{eq:reweight}
\end{aligned}
\end{equation}
where $Q(\beta)/Q(\beta_0)$ is computed via Eq.~\ref{eq:Q_ratio_quench}.

For a coordinate (possibly a vector) defined by a function $S(\textbf{x})$, the FES or potential of mean force (PMF) is given up to a constant factor by
\begin{equation}
    F(s,\beta) = -\frac{1}{\beta}\log\big(\left \langle \delta\big( S-s \big)\right\rangle\big),
\end{equation}
where $\delta$ is the Dirac delta function.\footnote{In real MD simulations, it is impossible to compute free energy at any particular $s$.
Rather, we use a block function (integrating delta function over windows), and we show in Appendix~\ref{app:C} that free energy computed in this way has an error with magnitude $\mathscr{O}({\Delta s}^2)$, where $\Delta s$ is the width of windows.}
Using Eq.~\ref{eq:reweight}, we then obtain our final result, which shows how the FES can be computed using quench trajectories,
\begin{equation}
     e^{-\beta F(s,\beta)}  \approx
    \frac{1}{N} \sum_{i=1}^N \frac{ \int_{\tau_i^{-}}^{\tau_i^{+}}dt \delta\big( S(\textbf{X}_i(t))-s \big) e^{-\beta \mathscr{H}(\mathbf{X}_i(t) )-d\gamma t} }
    {\int_{\tau_i^{-}}^{\tau_i^{+}} dt e^{-\beta_0 \mathscr{H}(\mathbf{X}_i(t))-d\gamma t}  }
    \label{eq:quenchpmf}
\end{equation}
This estimator allows us to compute the PMF at a range of temperatures $\beta$ above and below $\beta_0$ using a single set of trajectories.

 Because the exponential decay $e^{-d \gamma t}$ suppresses contributions at long forwards times and the exponential increase of $\mathscr{H}(\mathbf{X}_i(t))$ does so for large negatives times, we also considered running simulations for fixed forwards and backwards times where $\tau^+ \gg \tau_i^+(E_\mathrm{min})$ and $\tau^- \ll \tau_i^-(E_\mathrm{max})$. In this case,

\begin{equation}
     e^{-\beta F(s,\beta)}  \approx
    \frac{1}{N} \sum_{i=1}^N \frac{ \int_{\tau^{-}}^{\tau^{+}}dt \delta\big( S(\textbf{X}_i(t))-s \big) e^{-\beta \mathscr{H}(\mathbf{X}_i(t) )-d\gamma t} }
    {\int_{\tau^{-}}^{\tau^{+}} dt e^{-\beta_0 \mathscr{H}(\mathbf{X}_i(t))-d\gamma t}  },
    \label{eq:quenchpmf2}
\end{equation}
where now the integration limits are fixed for all runs. 
In Sec.~\ref{sec:results} we will give a heuristic for how long quench trajectories should be run.
In practice, to generate results we run fixed length simulations using that heuristic from many initial points, then pick energy cutoffs, and then use the estimator given by Eq.~\ref{eq:quenchpmf}, which did prove to be more accurate than Eq.~\ref{eq:quenchpmf2}.
This strategy allowed us to test both estimators and works well in practice, but does require more total simulation time than if a perfect energy cutoff were known \textit{a priori}.

Finally, we note that Ref.~\citenum{cao_learning_2022} proposes another exact estimator from the same type of trajectories, 
\begin{equation}
    \left \langle \phi \right \rangle \approx\frac{1}{N}\sum_{i=1}^{N}\int_{\tau^-}^{\tau^+} 
dt \frac{\phi({\bf X}_i(t)) \rho({\bf X}_i(t)) J(t) }{\int^{t-\tau^-}_{t-\tau^+}dt'\rho({\bf X}_i(t')) J(t')},
\label{eq:general_expression_cao}
\end{equation}
where $\tau^+$ and $\tau^-$ are constant. We also evaluate this formula for a test case in the Supporting Information, but find that it is more difficult to use in practice for molecular systems because (a) it requires obtaining data from ($\tau^-$-$\tau^+$, $\tau^+$-$\tau^-$), which is a strictly larger time window than in Eq.~\ref{eq:quenchpmf}, and (b) the long reverse quench to time $\tau^-$-$\tau^+$ can cause the MD simulation to become unstable as the kinetic energy grows exponentially, resulting in simulations crashing.
Hence we do not pursue it further in this work.

\section{IMPLEMENTATION}
We implemented quench dynamics in LAMMPS\cite{plimpton1995fast,thompson2022lammps} using the procedure described in Appendix \ref{sec:langevin}, with a user-defined ``fix'', and run trajectories using the LAMMPS python interface.
Then run many parallel trajectories in a Python framework using the parallel scripting language Parsl\cite{babuji2019parsl}, which also interfaces with common high-performance computing queuing systems.
We also perform analysis in parallel using parsl.
LAMMPS source code, as well as all run and analysis scripts, are provided in a github repository for this paper (\url{https://github.com/hocky-research-group/quench_paper_2023}).

\section{RESULTS}
\label{sec:results}
\subsection{Computing the partition function of independent harmonic springs through quenching}
To confirm the validity of the quench approach for a molecular system, as well as to check our implementation, we first start with a system for which we know the ground truth. 
We chose to study a system of $\mathscr{N}$ independent harmonic springs, with a Hamiltonian defined by,

\begin{equation}
    \mathscr{H}(\textbf{X})=\sum_{i=1}^{\mathscr{N}}{\left(\frac{|{\bf P}_i|^2}{2m}+\frac{1}{2}m\omega^2|{\bf Q}_i|^2\right)}
\label{eq:quench_EOMteen}
\end{equation}

Because these springs are independent, this is equivalent to $3\mathscr{N}$ one-dimensional harmonic oscillators defined by the simple Hamiltonian,

\begin{equation}
    \mathscr{H}(q,p)=\frac{p^2}{2m}+\frac{1}{2}m\omega^2q^2
\label{eq:harmonic_spring_1d}
\end{equation}

The partition function for this system is separable such that,
\begin{equation}
\begin{aligned}
    Q_{3\mathscr{N}}(\beta)&=Q^{3\mathscr{N}}_1(\beta) = \left( \int dq dp e^{- \frac{\beta p^2}{2m}- \frac{\beta}{2}m\omega^2q^2}\right)^{3\mathscr{N}} \\
                        &=\left( \sqrt{\frac{2\pi m}{\beta}}\times\sqrt{\frac{2\pi}{\beta m \omega^2}}\right)^{3\mathscr{N}} = \left({\frac{2\pi}{\beta \omega}}\right)^{3\mathscr{N}}
\end{aligned}
\end{equation}

We can therefore benchmark our quench approach by computing the ratio of partition functions at two different temperatures using Eq.~\ref{eq:Q_ratio_quench} and compare to the exact value, which is given by $(\beta_0/\beta)^{3\mathscr{N}}$.

Using LAMMPS, we sample $\mathscr{N}$ independent harmonic springs in 3D with identical masses $m=1.0$ and identical oscillation frequencies $\omega=\sqrt{5}$, in reduced units. We first generate 2000 starting points using Langevin dynamics \cite{jensen2019accurate} with friction coefficient $\gamma_{LD}=0.01$ and time step $\Delta t=0.001$ in reduced units.
To do so, we first equilibrate the system for $10^7$ steps ($\tau=10^4$ in reduced LJ time units) at $\beta=1$. Then we run production simulation for $2\times 10^7$ steps and save 2000 starting points for further ``quench'' simulations.

\begin{figure}[t!]
    \centering
    \includegraphics{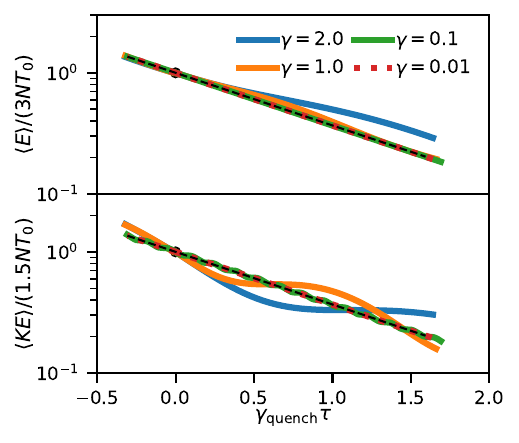}
    \caption{Mean total energy (top) and mean kinetic energy (bottom) with respect to reduced time ($\gamma_{quench}\tau$) for $\mathscr{N}=1000$ springs, varying $\gamma_\mathrm{quench}$, scaled by the equilibrium energy given by equipartition. Quenching is performed from 2000 starting points. Circles indicate the mean starting energies before quenching forwards and backwards in time.
    For small $\gamma_{quench}$, mean total energy and mean kinetic energy follow an exponential decay with time constant $\gamma^{-1}$ (dashed line).}
    \label{fig:mean_energy_gt}
\end{figure}

In Fig.~\ref{fig:mean_energy_gt}, we show the behavior of the energy of the system when running quench simulations using the EOM described by Eq.~({\ref{eq:quench_EOM}}) forward and backwards in time, for several values of $\gamma_\mathrm{quench}$.
We observe an overall exponential decay of average total energy at small $\gamma$ with respect to $\gamma_{quench} t$, a unitless ``time'' that we find serves as a good progress coordinate. 
In contrast, when $\gamma_{quench}$ is large, we observe large deviation from standard exponential decay, with a low-frequency oscillation.

To understand this behavior, we can solve the EOM of a 1d system exactly given an initial condition $(q_0,p_0)$,

\begin{equation}
\begin{aligned}
    &E(t)\exp{(\gamma t)}
    =\frac{2m^2\omega^4 q_0^2+2m\omega^2\gamma q_0p_0+2\omega^2 p_0^2}{m(4\omega^2-\gamma^2)}\\
    &+\frac{m^2\omega^2\gamma q_0^2-\gamma p_0^2}{2m\sqrt{4\omega^2-\gamma^2}}\sin{\left({\sqrt{4\omega^2-\gamma^2}\,t}\right)}\\
    &+\frac{m^2\omega^2\gamma^2q_0^2-4m\omega^2\gamma q_0p_0-\gamma^2p_0^2}{2(4\omega^2-\gamma^2)}\cos{\left( {\sqrt{4\omega^2-\gamma^2}\,t} \right)}
\end{aligned}
\label{eq:total_energy_expression}
\end{equation}

The magnitude of the ratio between terms scales for $\gamma<\omega$ as $\mathscr{O}(1):\mathscr{O}(\gamma/\omega):\mathscr{O}(\gamma^2/\omega^2)$ such that the first term dominates for small $\gamma$.
As $\gamma$ approaches $\omega$, then oscillations appear with a period of $\pi(\omega^2/\gamma^2-1/2)^{-1/2}$ when plotted against $\gamma t$.
Similarly, we can solve an expression for kinetic energy, and find that in the small $\gamma$ limit, the mean kinetic energy also decays exponentially with respect to time. 
In the quasi-static quenching limit, we can approximate all springs as independently following the same exponential decay, and therefore the sum of their energies also decays exponentially, 
 $E(t)\approx E(0) e^{-\gamma t}$. 
 For this situation, we can compute the partition function,
 \begin{equation}
\begin{aligned}
    Q(\beta)\propto&\int_{-\infty}^{\infty}{\exp{\left(-\beta E_0e^{-\gamma t}-d\gamma t\right)}\,dt}\\
    =&\int_{0}^{\infty}{\frac{1}{\gamma}\exp{\left(-\beta E_0 u\right)}u^{d-1}\,du}\\
    =&\frac{(d-1)!}{\gamma}(\beta E_0)^{-d}.
\end{aligned}
\label{eq:Q_beta}
\end{equation}
 This gives the correct result for the ratio of partition functions, $(\beta/\beta_0)^{-d}$.
Moreover, we could calculate $Q(\beta)$ using a saddle point approach and find that the exponential term is dominated by its value when $\gamma t=\ln{(\beta/\beta_0)}$, using the fact that $E_0 \approx d/\beta_0$ for Harmonic oscillators.

Because the mean kinetic energy decays exponentially with respect to time starting at a value of $d/(2\beta_0)$, at this particular moment $t=\gamma^{-1} \ln{(\beta/\beta_0)}$ the kinetic energy obtains a value of $d/(2 \beta)$ corresponding to a temperature $T$. 
In real simulations it is impossible to run to infinite times, so this harmonic model suggests that we can guess how long to run by choosing $\gamma(\tau^+-\tau^-) > \left|\ln{(\beta/\beta_0)}\right|$ for a target $\beta$ if we want to use the infinite time approximation as in Eq.~\ref{eq:quenchpmf2}.
To confirm this for the harmonic system, we show in Fig.~\ref{fig:Q_ratio_gt} that convergence of the ratio of partition functions reaches this limit where Eq.~\ref{eq:quenchpmf2} holds  once $\gamma (\tau^+-\tau^-)$ exceeds this value ($2\ln(2)\approx 1.4$). 
When $\gamma_\mathrm{quench}$ approaches $\omega$ and the exponential decay of energy does not hold, the partition function ratio converges to an incorrect value.

\begin{figure}[ht]
    \centering
\includegraphics{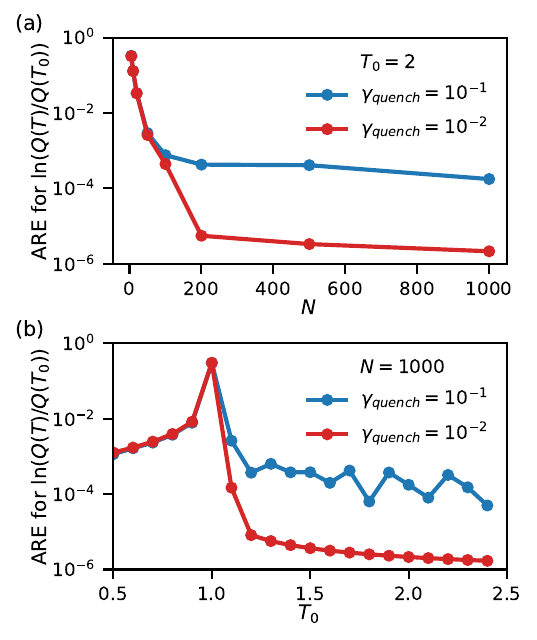}
    \caption{Absolute relative error in the ratio of partition functions at two different temperatures $T_0$ and $T=1$ for a system of independent harmonic springs using Eq.~\ref{eq:Q_ratio_quench}. (a) Error computed starting from $T_0=2$ and varying the number of springs for two different quench rates. (b) Error computed for fixed $N=1000$ springs when varying starting temperature at two different quench rates. Two other estimators are compared for this setup in Fig.~\ref{fig:compare_estimators_harmonic}. }
    \label{fig:harmonic_vary_N_T}
\end{figure}

Finally, we test the accuracy of our primary estimator Eq.~\ref{eq:Q_ratio_quench}. 
To do so, we take the same quenches performed for Fig.~\ref{fig:mean_energy_gt} and pick $\tau+$ and $\tau^-$ as the maximum final energy of the 2000 forward quenches and the minimum of the 2000 reverse quenches.
In Fig.~\ref{fig:harmonic_vary_N_T} we show that our quench estimator is highly accurate. 
Fig.~\ref{fig:harmonic_vary_N_T}a shows that the estimator gets more accurate with an increasing number of springs, despite the increasing phase space volume that must be sampled. Fig.~\ref{fig:harmonic_vary_N_T}b shows the effect of varying the initial temperature $T_0$, with more accurate results when $T_0$ is higher than $T$ as intuitively expected. 
Accuracy is improved substantially by decreasing the quench rate $\gamma$, but at the expense of longer simulations.

\subsection{Computing the FES for alanine through quenching}

\begin{figure*}[ht!]
    \centering
    \includegraphics{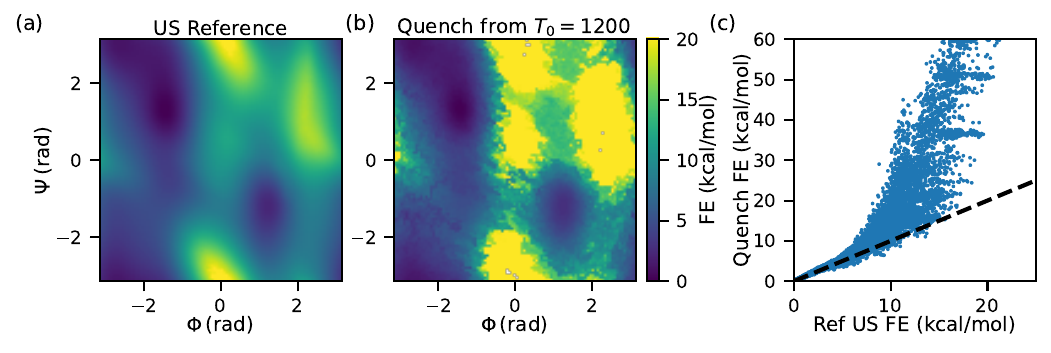}
    \caption{A comparison of FES between umbrella sampling and ``quench" method. (a) FES computed from US at $T=300$K with 400 ns total sampling time. (b) FES computed from ``quench" at $T=300$K with $T_0=1200$K with $\gamma_\mathrm{quench}=1\times10^{-4}$ ps$^{-1}$ and $10^4$ starting points corresponding to $\approx 410$ ns of total simulation time, of which data used in computing Eq.~\ref{eq:quenchpmf} (including 10 ns for generating restart points) totals $\approx 226$ ns.
    (c) Comparison of FES values on a bin-by-bin basis. While the minima are captured, the high free energy regions are not.}
    \label{fig:compare_ref_quench}
\end{figure*}
Having demonstrated that we are able to compute partition functions for a toy system, we are now interested in whether we can compute PMFs for a molecular system through a quenching procedure (where partition functions are not a very useful quantity on their own). 
We first test our approach by computing Eq.~\ref{eq:quenchpmf} for the test case of alanine dipeptide in vacuum, which has two major conformations (one of which has sub-populations) separated by a relatively large energy barrier, and whose configurations are well captured by considering its FES in the space of two backbone dihedral angles $\phi$ and $\psi$.
This FES has been extensively characterized by a number of enhanced sampling approaches and serves as a prototypical benchmark system, although one which is quite easy to sample for some approaches such as metadynamics.

Simulations of this molecule in vacuum were run using LAMMPS simulation package and using CHARMM 27 force field
without CMAP corrections\cite{mackerell2000development}. 
Equilibrium simulations are performed using LAMMPS's Langevin dynamics thermostat with $\gamma_\textrm{run}=0.01 \textrm{ps}^{-1}$ and an MD timestep of $\delta t = 1$ fs.
Umbrella sampling \cite{torrie1977nonphysical} was used to obtain reference free energy surface (FES). 
For umbrella sampling, a harmonic biased potential with spring constant 24.0 $\mathrm{kcal/(mol \cdot rad^2})$ was added to $400\ (20\times20)$ windows along CVs given by backbone dihedral angles $(\Phi,\Psi)$. The system was equilibrated for 400 ps at each window location, and then production runs were performed for 2 ns in each window. FES was estimated at the target temperature 300 K using WHAM to combine the data \cite{gallicchio2005temperature}, and was also computed for comparison using EMUS \cite{thiede2016eigenvector}. 
Finally, for comparison, FESs were computed using Well-Tempered MetaDynamics (WT-MetaD) \cite{barducci2008well} in PLUMED \cite{tribello2014plumed,plumed2019promoting} applied to $\phi,\psi$ with hills deposited every 500 steps, a hill width of 0.35 radians in each direction, a hill height of 0.286807 kcal/mol, and a bias factor of either 6 or 10 (see SI); WT-MetaD runs were performed for up to 100 ns although the FES estimate stopped changing within several nanoseconds. 
In ``quench" simulations, the system was equilibrated for 10 ns at $T_0=1200 K$ and 10000 starting points were drawn from a 10 ns production simulation with frequency every 1 ps. 
Fixed time simulations were performed with forward quenches of length $3.4 \gamma^{-1}$ and reverse quenches of length $0.6 \gamma^{-1}$. For this length forward quench, the final kinetic energy predicted by our exponential decay model is equivalent to approximately $T=110$ K, below where we want to estimate.
Times $\tau_i^+$ and $\tau_i^-$ were chosen from these data by histogramming the energies from the forward and reverse trajectories as shown in Fig.~\ref{fig:adp_energy_histogram}.

\begin{figure*}[ht!]
    \centering
    \includegraphics{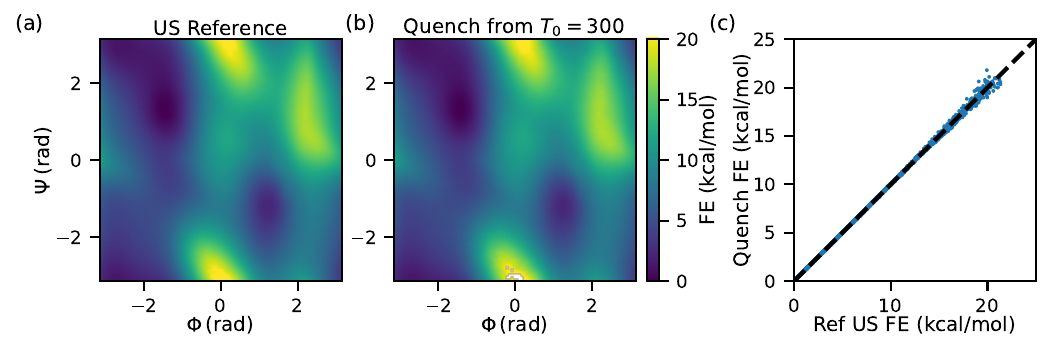}
    \caption{A comparison of FES between US and ``quench" combined with US. (a) FES computed from US at $T=300$K with 400 ns total sampling time after start point generation. (b) FES computed from by quench + US  sampling with 400 ns total sampling time including start point generation. Data used in computing Eq.~\ref{eq:quenchpmf} totals $\approx 287.6$ ns. (c) Comparison of FES values on a bin-by-bin basis. The quench+US landscape agrees almost exactly with the US one.}
    \label{fig:compare_ref_quench_umbrella}
\end{figure*}

Fig.~\ref{fig:compare_ref_quench} shows the comparison of FES computed by the ``quench" method with $\gamma_\mathrm{quench}=1\times10^{-4}$ ps$^{-1}$, for which the total amount of fixed time sampling was 400 ns, the same as that used in a reference US reference, and the time used with energy cutoffs was only 287 ns (see Tab.~\ref{tab:cost}).
While the shape of the FES was captured correctly using the ``quench" method, only the lower free energy regions were captured with high fidelity. As shown in Fig.~\ref{fig:compare_ref_quench}(c), quench results start to deviate from umbrella sampling results at approximately 8 kcal/mol$ \approx 13.5 k_B T$. 
Since this is a relatively high cutoff, it demonstrates that quench is applicable to molecular situations where the entire FES over some coordinates is not needed.
Moreover, with these simulations we should be able to estimate FES at any temperature in the range 200 K to 2200 K based on the amount of forward and reverse quenching performed, which would not be available with CV based approaches; this advantage is explored much more in the next section.
We also note that this was obtained in a CV agnostic manner.

We can also perform the calculations with larger $\gamma_\mathrm{quench}$ by a factor of 10, yielding slightly worse but comparable results (Fig.~\ref{fig:adp_fast_quench}, showing that the important regions of the landscape can be captured efficiently in tens of nanoseconds, which is similar to the convergence speed of WT-MetaD \cite{barducci2008well} (see also next section).
We can moreover perform this calculation with a 10 or 100 times slower quench, resulting in much more accurate results, but with a relatively small improvement compared to the amount of additional sampling (Fig.~\ref{fig:adp_slow_quench},~\ref{fig:adp_extraslow_quench}).
Next, we show that if we wish to resolve the high free energy regions in detail, it is possible to combine quenching and CV based approaches.

\begin{figure*}[ht!]
    \centering
    \includegraphics{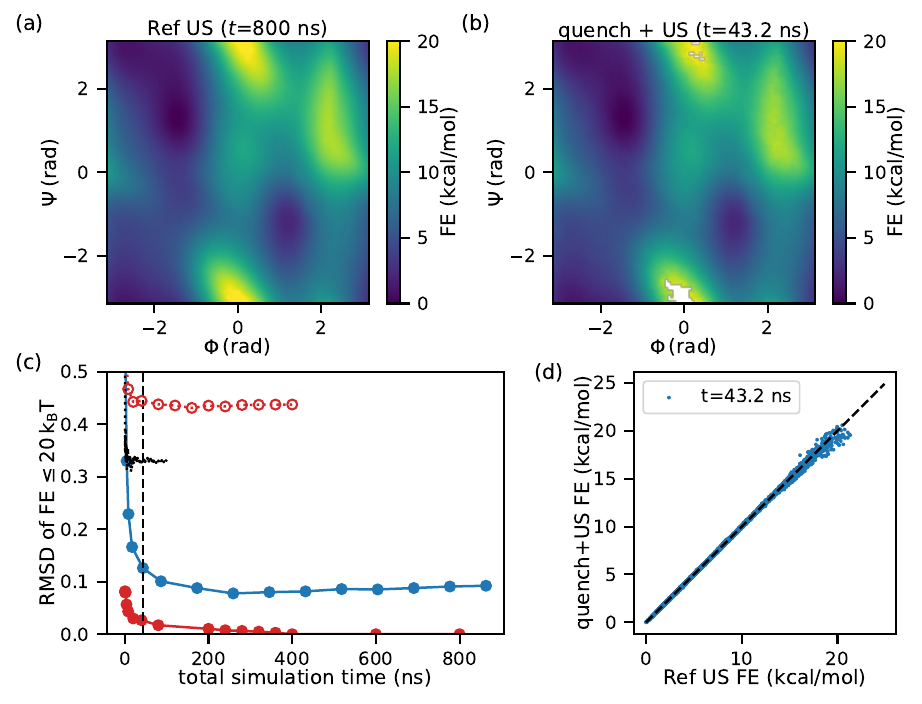}
    \caption{(a) Reference US FES with 800 ns of simulation. (b) FES computed from quench + US using 50 quenches per window corresponding to 60 ns of total simulation data, with $~\sim 41$ ns used to compute Eq.~\ref{eq:quenchpmf}. (c) RMSD of free energy for bins with FE  $\leq 20\ k_BT$ (in kcal/mol) compared to reference umbrella sampling as a function of total sampling time, which is adjusted by changing the number of starting points used. Black dashed line shows comparison with WT-MetaD using bias factor 6 (see Fig.~\ref{fig:metad_bf6}), and open red circles show error when computing FES on reference data using EMUS rather than WHAM (see Fig.~\ref{fig:emus_si_2}). (d) Comparison of FES values in (a) and (b) on a bin-by-bin basis.}
    \label{fig:adv1}
\end{figure*}

 \subsection{Combining quenching and umbrella sampling}
Because quenching does a good job capturing the low free energy regions, we predicted that it could be combined with US to produce an efficient sampling approach. 
By nature of introducing a harmonic potential, we obtain by construction a more convex landscape, albeit one that can still be hard to sample due to slow orthogonal degrees of freedom.
To directly compute the FES from our quench data, we also derived a modified version of the WHAM equations following the derivation of Ref.\citenum{gallicchio2005temperature} (See Appendix~\ref{app:D}). Surprisingly, it turns out that we can use exactly the same WHAM equations when $\beta=\beta_0$ by realizing that Jacobian $J(t)$ is same among umbrella windows and the ratio of any two umbrella windows are fixed over time.
When we do not start from the target temperature, we can estimate the biased densities in from each window and combine them with the usual WHAM equations. \cite{roux1995calculation}
In Sec.~\ref{sec:emus} we also show that we can apply the Eigenvector Method for Umbrella Sampling \cite{thiede2016eigenvector} (EMUS) with our quench data.
EMUS is a meshless estimator like MBAR and other post-WHAM approaches \cite{thiede2016eigenvector,shirts2008statistically,ding2019fast}.
In other words, it uses the true bias potentials instead of approximate ones to estimate the weight of a sample generated in one umbrella versus that in another. 
EMUS is more expensive to apply than WHAM and we did not find the results to be significantly different, hence we demonstrate that it works in the Supporting Information but include results from WHAM in the main text. 

We first tested combining these methods by taking starting points generated by the equilibrium US procedure at $T=300$K, and quench in the presence of the same harmonic bias with $\gamma_\textrm{quench}=0.001$ ps$^{-1}$ for a forward time of $1 \gamma^{-1}$ and a reverse time of $1 \gamma^{-1}$, with the idea that we can use our estimator to compute unbiased weighted histograms in each position and then combine the results with any standard free energy approach. 
We note that this quench time can be thought of as approximately increasing and decreasing the temperature by a factor of $e$, giving access to temperatures between approximately 110 and 815 K.
For each window, we select an upper and lower bound to fix $\tau_i^{+}$ and $\tau_i^{-}$ as in the previous section.

Fig.~\ref{fig:compare_ref_quench_umbrella} shows the comparison of FES between US and quench+US.
The resulting FES is almost identical between our reference result and our newly computed surface up to the maximum range accessed from US.
We also show that this is the case when the quench FES is computed using EMUS, when comparing to a reference FES computed by EMUS or by WHAM in Fig.~\ref{fig:emus_si_1} and ~\ref{fig:emus_si_2}, respectively.

Although so far we have reported results for a total equivalent amount of sampling time of 400 ns (with 287.6 ns being used in the estimator), the results of quench+US can be obtained much more quickly than that.
In  Fig.~\ref{fig:adv1} we show that the full FES converges much more quickly than that. In Fig.~\ref{fig:adv1}b we highlight the case of 60 ns (43.2 ns used for the estimator) at which point there is virtually no error and only some very high energy regions are not fully sampled.  Fig.~\ref{fig:adv1}c shows that the error is already  minimal by $\sim 10$ ns and stops decreasing by $~\sim 200$ ns.
Although the deviation between US and quench+US does not converge to zero, we also show in this plot that this difference is much smaller than that obtained when using a different method (WT-MetaD) or even a different method of estimating the FES from the same reference data (EMUS, open red circles). 
Thus we do not consider an error of 0.1 kcal/mol over the entire surface to be significant.

\begin{figure}
    \centering    \includegraphics{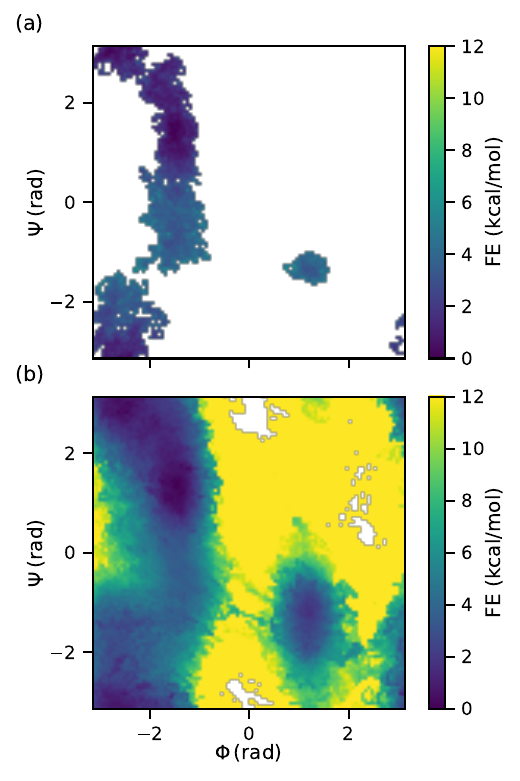}
    \caption{FES computed using only one bad CV. (a) FES computed from US on $\Psi$ with 40 ns total sampling. (b) FES computed from quench+US from $T_0=1200$ on $\Psi$ with approximately 40 ns total sampling (28.5 ns used in computing Eq.~\ref{eq:quenchpmf}), showing much more exploration. }
    \label{fig:adv2}
\end{figure}

Next, we show that quenching helps US when using a bad CV.
This is demonstrated by considering the case of US only along $\psi$, which does not distinguish the positive and negative $\phi$ basins well. 
Two sets of simulations using the same US parameters were run, biasing at 20 windows along $\Psi$ using the same spring constant of 24.0 $\mathrm{kcal/(mol \cdot rad^2})$. 
The reference US simulation used 2 ns per window, resulting in 40 total ns of simulation.
For quenching, we use 666 starting points separated by 1 ps in each window, with $\gamma_\textrm{quench}=0.001$  ps$^{-1}$ for $\gamma \tau^+=1,\gamma \tau^-=-1$ corresponding to an equivalent amount of simulation time and 28.5 ns used in the estimator.
The FES computed from US and WHAM (Fig.~\ref{fig:adv2}a) shows that little sampling is achieved.
When adding quenching, the heating phase allows the system to overcome some hidden energy barriers that are not captured by bad CV $\Psi$, resulting in a surface which captures all minima relatively well but does not resolve the barrier between the basins correctly.

\begin{figure*}
    \centering
    \includegraphics{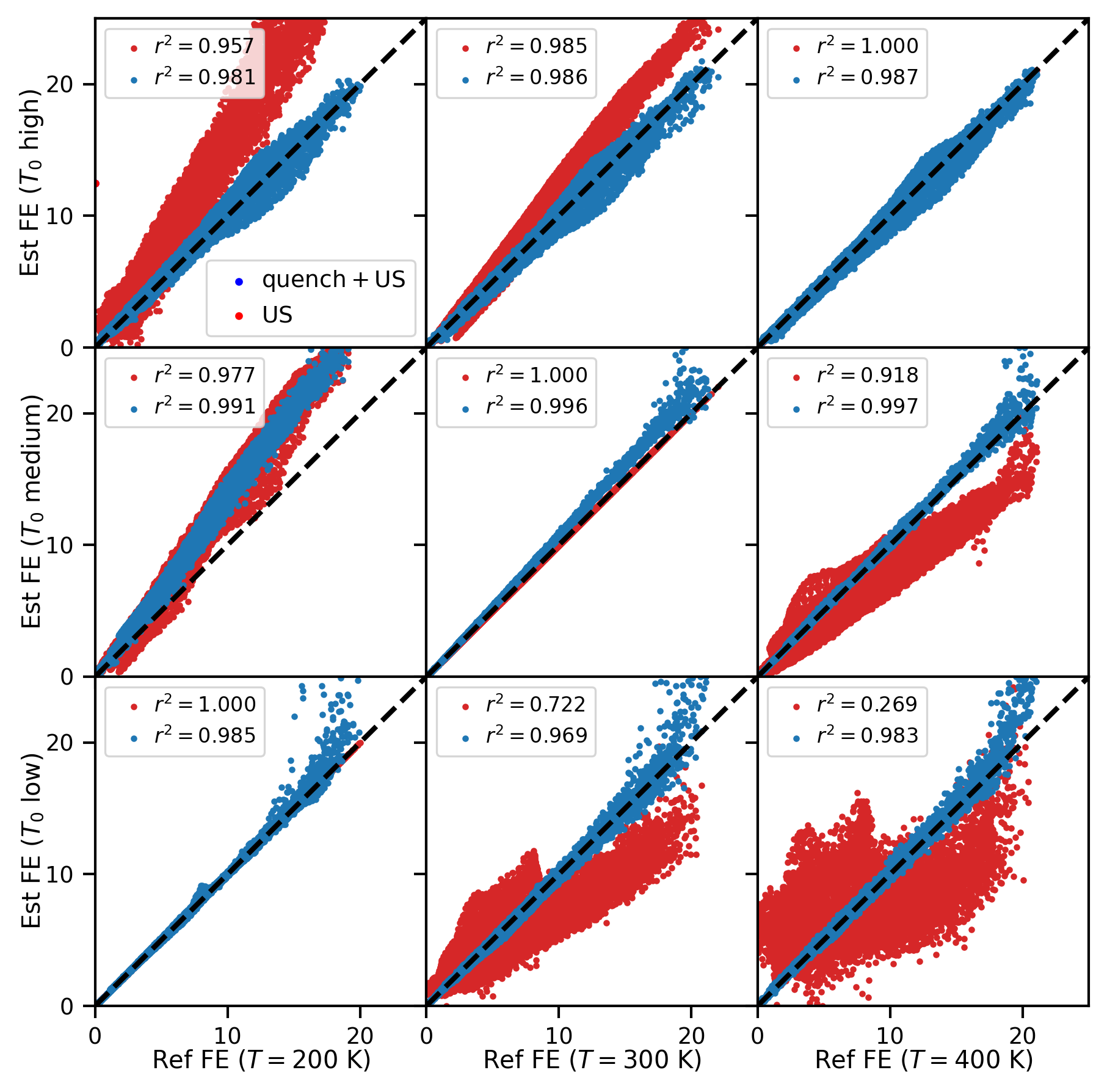}
    \caption{ Comparison of the quality of FES computed by sampling at $T_0$ and estimating at $T$. In each case, the $x$-axis shows the FES computed by 800 ns of US at a reference temperature of either $T=$ 200, 300, or 400 K. The $y$-axis shows the FES computed at $T$ when starting at a low, medium, or high $T_0$. For US (red circles), these are $T_0=$200, 300, 400 K and for quench+US (blue circles) these are 75, 300, and 1200 K. US alone fails at extrapolating even by 100 K (33\%) while quench+US is much more robust. For US, total simulation time is 800 ns and for quench+US, $\gamma_\mathrm{quench}=0.001$  ps$^{-1}$ and total sampling time is 800 ns (see Tab.~\ref{tab:cost}). To be consistent, the FES for quench+US is computed in all cases by estimating the unbiased density in each case and combining by WHAM, even though we could use our exact WHAM equation for the $T=T_0=300$ case. }
    \label{fig:adv3}
\end{figure*}

Last, we again emphasize that quenching in principle allows estimating FES at various target temperatures using the same simulation data.
To quantify this, we tested whether we could obtain the FES at a range of temperatures, quenching from above, below, and in the middle. 
Using starting points drawn from $T_0=$ 75, 300, and 1200K, and three quench simulations, we demonstrate that the FES can be obtained for a wide range of temperatures above and below 300 K. 
In Fig.~\ref{fig:adv3} we show the bin-by-bin comparison with US for target temperatures $T=$ 200, 300, and 400 K.
In all but one case, the results are quite robust across the whole FES.
In contrast, US reweighted to other temperatures using the WHAM equations performs poorly in all six cases tested despite using twice the amount of total sampling data. 
The one case shown where quench+US fails is sampling initial points from $T_0=$300 K and estimating at $T=$200 K, which may be due to insufficient sampling in the chosen initial points at $T_0=$300 K that was not evident when estimating at higher temperatures. 

\section{PRELIMINARY EXTENSION TO SOLVATED SYSTEMS}
\begin{figure*}
    \centering
    \includegraphics{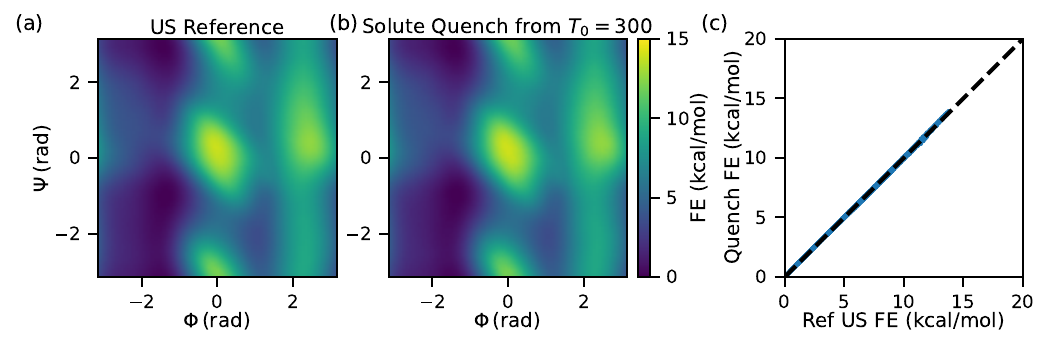}
    \caption{A comparison of FES of alanine dipeptide in water between US and solute quench+US. (A) FES computed from umbrella sampling at $T=$300 K  using 800 ns of total data. (B) FES computed from s-quench+US with $T_0=300$ K using 400 ns of total data. (c) Bin-by-bin comparison of FES between these two cases shows exact agreement.}
    \label{fig:alanine_solv_compare}
\end{figure*}

Most MD simulations are run in solution which adds thousands of additional DOF from solvent molecules. 
We were concerned that quenching could present difficulties in this case both (a) due to numerical issues arising from the extensivity of the terms in the exponential within our FES estimation formula, and (b) due to the problem of super-heating the solution.
To mitigate these issues, we propose a ``solute-quench'' approach (s-quench) in the spirit of solute tempering \cite{liu2005replica}.
Here we investigate a scheme where the solvent degrees of freedom are evolved by Newtonian dynamics and the solute degrees of freedom follow quench dynamics. 
Mathematically, this is expressed as,

\begin{equation}
    \left \{ 
    \begin{aligned}
    \dot{{\bf Q}}_{\rm{solute}} & = {\bf M}_{\rm{solute}}^{-1}{\bf P}_{\rm{solute}} \\
    \dot{{\bf P}}_{\rm{solute}} & = -\boldsymbol{\nabla}_{\rm{solute}}{U({\bf Q}_{\rm{all}})}-\gamma {\bf P}_{\rm{solute}}\\
    \dot{{\bf Q}}_{\rm{solvent}} & = {\bf M}_{\rm{solvent}}^{-1}{\bf P}_{\rm{solvent}} \\
    \dot{{\bf P}}_{\rm{solvent}} & = -\boldsymbol{\nabla}_{\rm{solvent}}
    U({\bf Q}_{\rm{all}})
    \end{aligned} \right .
\label{eq:quench_solv_EOM}
\end{equation}
For these equations, the Jacobian is still trivial to derive,

\begin{equation}
    J(t)=\exp{(-d_{\rm{solute}}\gamma t)}
\label{eq:quench_solv_J}
\end{equation}

It is possible to implement this algorithm in LAMMPS with same ``fix'' by simply choosing two groups of atoms to integrate separately by two different integrators.
We tested s-quench on alanine dipeptide in water.
Using CHARMM-GUI \cite{jo2008charmm,lee2016charmm,lee2020charmm} we generated a LAMMPS input file for alanine in a box of dimension 28\AA\ per side with 752 water molecules. 
As in vacuum, the system is equilibrated for 200 ps in each window, and then the rest of the US protocol is the same.
Here, the forcefield used is CHARMM36m.
We show in Fig.~\ref{fig:alanine_solv_compare} that this approach works and s-quench+US gives identical results to standard US in this case.
Interestingly, due to the small size of the solute relative to the solvent, energy fluctuations make the histograms of minimum and maximum energy overlap. 
Hence for this case we rely on the infinite time approximation of Eq.~\ref{eq:quenchpmf2}.

\section{CONCLUSIONS}
In the system of independent harmonic springs, a nearly exponential decay of the mean total energy and the mean kinetic energy has been observed in the limit of small $\gamma_{quench}$. With this observation, we can use $E(t)=E(0)e^{-\gamma t}$ to predict the results of Eqn.~\ref{eq:quench_expression} or Eqn.~\ref{eq:reweight}. We found that these results give a good heuristic for how long to run quench simulations in practice.

Quenching alone gave moderate performance for alanine in vacuum if the desired quantity is a fast estimation of the entire FES at a single temperature, and would not be the method of choice at least for this simple problem where e.g. metadynamics is extremely efficient as demonstrated in many papers. 
On the other hand, quenching does give access to the FES at a range of temperatures in a single shot, and so it may be efficient in cases where that is desired.
Moreover, we demonstrated that quench combined with US is accurate and efficient across a range of temperatures.
Quench+US can also alleviate issues in US due to hidden slow unbiased CVs, since reverse quenching heats up the system and high free energy regions are more likely to be sampled.

Here, we demonstrated that the WHAM equations also apply to this method when starting temperature $T_0$ is equal to target temperature $T$, and showed that it is possible to estimate the FES at other temperatures using a density-based WHAM approach. 
It is known that WHAM is not the most optimal method of reweighting data, and other approaches could be used to combine quench+US data; e.g. we tried combining US + quench data using EMUS \cite{thiede2016eigenvector}, but this was more computationally demanding and resulted in larger errors due to numerics.

Furthermore, we proposed a solute tempering version of quenching so that it can be applied to solution systems where there are many DOF due to the presence of solvent molecules. 
We showed that a perfect FES can also be obtained by combining this approach with US.
As with solute tempering, the major advantage of estimating the FES at other values of $T$ is no longer applicable, and so it remains to be tested whether this approach is more efficient than the equilibrium approach of US combined with solute tempering for a more difficult system.

Given our results, we feel that estimators based on nonequilibrium trajectories can offer an alternative approach worth considering. 
These approaches may prove to be well suited to certain classes of problems that we have not yet tested, such as computing basin volumes or phase equilibria of simple systems, and we plan to do so going forward.

\section{Data availability}
All code and input scripts used to generate data for figures, as well as the data used to produce the figures are available from \url{https://github.com/hocky-research-group/quench_paper_2023}.

\begin{acknowledgments}
KL and GMH were supported by the National Institutes of Health through the award R35GM138312. KL was also partially supported by a graduate fellowship from the Simons Center for Computational Physical Chemistry (SCCPC) at NYU (SF Grant No. 839534), and earlier by a grant from the Department of Energy (DE-SC0020464). 
GMR was supported by the U.S. Department of Energy, Office of Science, Office of Basic Energy Sciences, under Award Number DE-SC0022917.
EVE was supported by the National Science Foundation under Awards DMR-1420073, DMS-2012510, and DMS-2134216, by the Simons Collaboration on Wave Turbulence, Grant No. 617006, and by a Vannevar Bush Faculty Fellowship.
This work was supported in part through the NYU IT High Performance Computing resources, services, and staff expertise, and simulations were partially executed on resources supported by the SCCPC at NYU.
\end{acknowledgments}

\appendix

\section{Derivation of the Jacobian $J(t)$}
\label{sec:phasevolume}

The Jacobian $J(t)$ can be understood as a time-dependent factor that describes how phase space volume changes over time because of the non-equilibrium process. We can compute the change in phase space volume analogous to how it is done for the derivation of Liouville's theorem \cite{kardar2007statistical}. 
First, let us consider an infinitesimal phase space volume at time $t$, defined by $d+1$ arbitrarily closely spaced points $\mathbf{X}(t)=(X_1,...,X_d),\left\{\mathbf{X}'_i(t)=\mathbf{X}(t)+\delta\mathbf{X}_i(t)\right\}_{i=1,...,d}$. An infinitesimal phase space volume at time $t$ is defined as follows,

\begin{equation}
    \delta V(t) = \det{\left(\delta{\bf X}_1(t),...,\delta{\bf X}_d(t)\right)},
\end{equation}
where $\delta{\bf X}_i(t) = {\bf X}'_i(t) - {\bf X}(t)$.

After infinitesimal time $\delta t$,

\begin{widetext}
\begin{equation}
    \left \{ \begin{aligned}
    {\bf X}_i(t+\delta t) &=  {\bf X}_i(t) + \boldsymbol{b}({\bf X}(t))\,\delta t+\mathscr{O}(\delta t^2)\\
    {\bf X}'_i(t+\delta t) &=  {\bf X}'_i(t) + \boldsymbol{b}({\bf X}'_i(t))\,\delta
    t+\mathscr{O}(\delta t^2)\\
    &={\bf X}_i(t)+\delta {\bf X}_i(t)+\left[\boldsymbol{b}({\bf X}(t))+\boldsymbol{\nabla}\boldsymbol{b}({\bf X}(t))\cdot\delta{\bf X}_i(t)+\mathscr{O}\left(|\delta {\bf X}_i|^2(t)\right)\right]\delta t+\mathscr{O}(\delta t^2)\\
    \delta {\bf X}_i(t+\delta t) &= {\bf X}'_i(t+\delta t)-{\bf X}_i(t+\delta t)\\
    &\approx \delta {\bf X}_i(t)+\left[{\boldsymbol{\nabla}\boldsymbol{b}({\bf X}(t))\cdot\delta{\bf X}_i(t)}\right]\delta t+\mathscr{O}(\delta t^2)\\
    &= \left({\bf I}+\delta t\boldsymbol{\nabla}\boldsymbol{b}({\bf X}(t))\right)\cdot\delta{\bf X}_i(t)+\mathscr{O}(\delta t^2)\\
    {\bf I} + \delta t\boldsymbol{\nabla}\boldsymbol{b}({\bf X}(t)) &=
    \begin{pmatrix}
    1 + \delta t\frac{\partial b_1({\bf X}(t))}{\partial X_1} & \hdots & \delta t\frac{\partial b_1({\bf X}(t))}{\partial X_d}\\
    \vdots &  & \vdots \\
    \delta t\frac{\partial b_d({\bf X}(t))}{\partial X_1} & \hdots & 1 + \delta t\frac{\partial b_d({\bf X}(t))}{\partial X_d}
    \end{pmatrix}\\
    \delta V(t+\delta t) &= \det{\left({\delta{\bf X}_1(t+\delta t),...,\delta{\bf X}_d(t+\delta t)}\right)}\\
    &\approx \det{\left({\left({\bf I}+\delta t\boldsymbol{\nabla}\boldsymbol{b}({\bf X}(t))\right)\cdot\left({\delta{\bf X}_1(t),\hdots,\delta{\bf X}_d(t)}\right)}\right)}\\
    &= \det{\left({\bf I}+\delta t\boldsymbol{\nabla}\boldsymbol{b}({\bf X}(t))\right)}\cdot\det{\left(\delta{\bf X}_1(t),\hdots,\delta{\bf X}_d(t)\right)}\\
    &= \left(1+\delta t\boldsymbol{\nabla} \cdot \boldsymbol{b}({\bf X}(t)) + \mathscr{O}(\delta t^2)\right)\delta V(t)
    \end{aligned} \right.
\end{equation}
\end{widetext}

The last equation gives at lowest order,

\begin{equation}
    \frac{1}{\delta V(t)}\frac{d}{dt}\delta V(t) = \boldsymbol{\nabla} \cdot \boldsymbol{b}({\bf X}(t))
\end{equation}

and by integrating this equation gives

\begin{equation}
\begin{aligned}
    \delta V(t) &= \delta V(0) \times \exp{\left(\int_0^t{\boldsymbol{\nabla} \cdot \boldsymbol{b}({\bf X}(s)) \,ds}\right)}\\
    &= \delta V(0) \times J(t)
\end{aligned}
\end{equation}

\section{Time evolution of quench equations of motion}
\label{sec:langevin}
Since the functional form of quench is similar to Langevin Dynamics, we use the so-called BAOAB scheme \cite{leimkuhler2013robust} to get an accurate numerical update of coordinates and momenta. We split quench equation of motion into three parts.

\begin{itemize}
    \item (B) is $\dot{\bf P}(t) = -{\boldsymbol{\nabla}}U\left({\bf Q}(t)\right)$
    \item (A) is $\dot{\bf Q}(t) = {\bf M}^{-1}{\bf P}(t)$
    \item (O) is $\dot{\bf P}(t) = -\gamma {\bf P}(t)$
\end{itemize}

By iterating these parts in the order BAOA for discrete time steps proportional to $\delta t$, time is advanced.
We can solve the (O) step analytically to advance time more accurately, while the other two steps are advanced to first order in $\delta t$.
This results in a scheme:

\begin{equation*}
    \left \{ \begin{aligned}
    \textrm{(B)}: {\bf P} &\leftarrow {\bf P} - {\boldsymbol{\nabla}}U\left({\bf Q}\right)\cdot\delta t\\
    \textrm{(A)}: {\bf Q} &\leftarrow {\bf Q} + {\bf M}^{-1}{\bf P}\cdot\frac{\delta t}{2}\\
    \textrm{(O)}: {\bf P} &\leftarrow \exp{\left(-\gamma\delta t\right)}{\bf P}\\
    \textrm{(A)}: {\bf Q} &\leftarrow {\bf Q} + {\bf M}^{-1}{\bf P}\cdot\frac{\delta t}{2}
    \end{aligned} \right.
\end{equation*}

This is implemented in a LAMMPS `fix' \texttt{quench\_exponential}, implemented by the code ``fix\_quench\_exponential.cpp'' and ``fix\_quench\_exponential.h'' available in the github repository for this paper.

\section{Numerical error in free energy calculation}
\label{app:C}

We compute free energy surfaces on a discrete grid, resulting in a discretization error. 
First, let us consider free energy along 1D CV $s\in\mathbb{R}$, but this can be generalized to higher dimensional CVs in a similar manner. We assume the true free energy $F(s)$ is differentiable.

\begin{equation}
    F(s)=-\frac{1}{\beta}\ln{\left(\frac{\int_{\mathbb{R}^{2d}}{\delta (S({\bf X})-s)\exp{\left(-\beta\mathscr{H}({\bf X})\right)}\,d{\bf X}}}{\int_{\mathbb{R}^{2d}}{\exp{\left(-\beta\mathscr{H}({\bf X})\right)}\,d{\bf X}}}\right)}
\label{eq:B1}
\end{equation}

In practice, this delta function is replaced by a window function which is 1 within a range of s and zero outside. 

Suppose $\bar{F}(s_0)$ is the average free energy we should get at the specific site $s_0$ over an interval with length $\Delta s$.

\begin{equation}
\begin{aligned}
    \bar{F}(s_0)\Delta s &= \int_{s_0-\Delta s/2}^{s_0+\Delta s/2}{F(s)\,ds}\\
    &= \int_{-\Delta s/2}^{\Delta s/2}{F(s_0+s)\,ds}\\
    &= \int_{-\Delta s/2}^{\Delta s/2}{F(s_0)+F'(s_0)s+\frac{1}{2}F''(s_0)s^2+\mathscr{O}(s^3)\,ds}\\
    &= F(s_0)\Delta s+\frac{1}{24}F''(s_0)\Delta s^3+\mathscr{O}(\Delta s^5)
\end{aligned}
\label{eq:B2}
\end{equation}

To apply this averaging idea, we can apply it to the probability distribution rather than the free energy itself, resulting in:

\begin{widetext}
\begin{equation}
\begin{aligned}
    \exp{\left(-\beta\tilde{F}(s_0)\right)}\Delta s &= \int_{s_0-\Delta s/2}^{s_0+\Delta s/2}{\exp{\left(-\beta F(s)\right)}\,ds}\\
    &= \int_{-\Delta s/2}^{\Delta s/2}{\exp{\left(-\beta F(s_0+s)\right)}\,ds}\\
    &= \int_{-\Delta s/2}^{\Delta s/2}{\exp{\left(-\beta\left[{F(s_0)+F'(s_0)s+\frac{1}{2}F''(s_0)s^2+\mathscr{O}(s^3)}\right]\right)}\,ds}\\
    &\approx \exp{\left(-\beta F(s_0)\right)}\int_{-\Delta s/2}^{\Delta s/2}{1-\beta F'(s_0)s+\frac{1}{2}\left({\beta^2{F'}^2(s_0)-\beta F''(s_0)}\right)s^2+\mathscr{O}(s^3)\,ds}\\
    &= \exp{\left(-\beta F(s_0)\right)}\left[{\Delta s+\frac{1}{24}\left({\beta^2{F'}^2(s_0)-\beta F''(s_0)}\right)\Delta s^3+\mathscr{O}(\Delta s^5)}\right]
\end{aligned}
\label{eq:B3}
\end{equation}
\end{widetext}

From Eq.~(\ref{eq:B2}) and Eq.~(\ref{eq:B3}), we find that the true free energy, the mean free energy over an interval, and the free energy computed from the block functions are approximately equal within first order accuracy and the error has a magnitude of $\mathscr{O}(\Delta s^2)$.

\section{Derivation of WHAM equations}
\label{app:D}

WHAM\cite{kumar1992weighted} (Weighted Histogram Analysis Method) is a widely-used technique to reweight data from different windows in umbrella sampling simulations. 
However, WHAM is valid only for an equilibrium process, so here we derive a nonequilibrium version of WHAM that applies in this case.
We follow the derivation in Ref.~\citenum{gallicchio2005temperature}, and modify some parts to match the ``quenching'' case.

In the case of regular umbrella sampling simulations, let $\rho_{kl}^{\circ}$ be the unbiased probability density at window $(\phi_k,\psi_l)$ which we wish to determine. 
To compute this, we apply bias potentials $\omega_{ij,kl}$ at a window $(\phi_k,\psi_l)$ with an additional harmonic potential centered at site $(\phi_i,\psi_j)$ defined by:

\begin{equation}
    \omega_{ij,kl} = \frac{1}{2}\kappa\left[{(\phi_i-\phi_k)^2+(\psi_j-\psi_l)^2}\right]
\end{equation}
Note that because the bias is applied to dihedral angles, dihedral differences are computed taking into account periodicity of $2\pi$ radians.
Under the influence of this potential, we measure $n_{ij,kl}$ the `counts' (number of sampled data points) in the window centered at $(\phi_k,\psi_l)$.

Since the ratio of probability densities between two windows is fixed, the overall probability density $\rho_{ij,kl}$ at window $(\phi_k,\psi_l)$ with biased potential at site $(\phi_i,\psi_j)$ is the linear combination of biased probability densities:

\begin{equation}
    \rho_{ij,kl} = c_{ij}\rho_{kl}^{\circ}\exp{\left(-\beta_0\omega_{ij,kl}\right)},
\end{equation}

where $c_{ij}$ is the normalization factor at $t=0$:

\begin{equation}
    c_{ij}^{-1} = \sum_{k,l}{\rho_{kl}^{\circ}\exp{\left(-\beta_0\omega_{ij,kl}\right)}}
\end{equation}

The overall probability of getting the sampled data is proportional to the product of these biased probability densities:

\begin{equation}
\begin{aligned}
    \mathbb{P} &\propto \prod_{i,j}{\prod_{k,l}{(\rho_{ij,kl})^{n_{ij,kl}}}}\\
    &= \prod_{i,j}{\prod_{k,l}{\left(c_{ij}\rho_{kl}^{\circ}\exp{\left(-\beta_0\omega_{ij,kl}\right)}\right)^{n_{ij,kl}}}}
\end{aligned}
\end{equation}

We can estimate the true probability density by maximizing the log-likelihood, 

\begin{equation}
\begin{aligned}
    \frac{\partial\ln{\mathbb{P}}}{\partial\rho_{kl}^{\circ}} &= \sum_{i,j}{n_{ij,kl}\frac{1}{\rho_{kl}^{\circ}}+\sum_{i,j}{\sum_{k,l}{n_{ij,kl}\frac{1}{c_{ij}}\frac{\partial c_{ij}}{\partial\rho_{kl}^{\circ}}}}}\\
    &= \sum_{i,j}{n_{ij,kl}\frac{1}{\rho_{kl}^{\circ}}-\sum_{i,j}{\left({\sum_{k,l}{n_{ij,kl}}}\right)c_{ij}\exp{\left(-\beta_0\omega_{ij,kl}\right)}}}\\
    &= 0
\end{aligned}
\end{equation}

Summarizing this equation and normalization condition gives WHAM equations for regular umbrella sampling simulations:

\begin{equation}
    \left \{ \begin{aligned}
    \rho_{kl}^{\circ} &= \frac{\sum_{i,j}{n_{ij,kl}}}{\sum_{i,j}{\left({\sum_{k,l}{n_{ij,kl}}}\right)c_{ij}\exp{\left(-\beta_0\omega_{ij,kl}\right)}}}\\
    c_{ij}^{-1} &= \sum_{k,l}{\rho_{kl}^{\circ}\exp{\left(-\beta_0\omega_{ij,kl}\right)}}
    \end{aligned} \right.
\label{eq:C6}
\end{equation}

Similarly, in quenching, let $\rho_{ij,kl,t}$ be probability density at window $(\phi_k,\psi_l)$ and at time $t$ with biased potential centered at $(\phi_i\psi_j)$:

\begin{equation}
    \rho_{ij,kl,t} = c_{ij}\rho_{kl}^{\circ}\exp{\left(-\beta_0\omega_{ij,kl}\right)}\exp{(-d\gamma t)},
\end{equation}

where $c_{ij}$ is the normalization factor:

\begin{equation}
    c_{ij}^{-1} = \sum_{k,l}{\rho_{kl}^{\circ}\exp{\left(-\beta_0\omega_{ij,kl}\right)}}
\end{equation}

We eventually get exactly the same equation as Eq.~(\ref{eq:C6}) except for the expression of $n_{ij,kl}$:

\begin{equation}
    n_{ij,kl} = \sum_{t}{n_{ij,kl,t}}
\end{equation}

Note that $\rho_{kl}^{\circ}$ is unbiased probability density at starting temperature $\beta_0$. If we would like to estimate the unbiased probability $\rho_{kl}^{\circ}$ at the target temperature $\beta$ while the simulations are run at a different starting temperature $\beta_0$, then there are two ways to compute $n_{ij,kl}$.\par
In the first approach, we can estimate the density of states $n(\phi_k,\psi_l,E)$ instead with the following relations:

\begin{equation}
    \begin{aligned}
        \exp{\left(-\beta F(\phi_k,\psi_l)\right)} &\approx \sum_{E}{n(\phi_k,\psi_l,E)\exp{\left(-\beta E\right)}\Delta E}\\
    &\approx \sum_{E}{n(\phi_k,\psi_l,E)\exp{\left(-\beta E\right)}E\Delta \ln{E}}
    \end{aligned}
\end{equation}
Here the conversion to use $\Delta \ln({E})$ is shown because, in practice, the exponential decay/increase in energy from quench dynamics leads to a very wide range of energy values, and hence it is more computationally convenient to histogram the log of the energy. 

\begin{equation}
    \left \{ \begin{aligned}
    n(\phi_k,\psi_l,E) &= \frac{\sum_{i,j}{n_{ij,kl,E}}}{\sum_{i,j}{\left(\sum_{k,l,E}{n_{ij,kl,E}}\right)c_{ij}\exp{\left(-\beta_0\left(E+\omega_{ij,kl}\right)\right)}}}\\
    c_{ij}^{-1} &= \sum_{k,l,E}{n(\phi_k,\psi_l,E)\exp{\left(-\beta_0\left(E+\omega_{ij,kl}\right)\right)}}
    \end{aligned} \right.
\end{equation}

In the second, we compute ``effective counts''. We use Eq.~(\ref{eq:quenchpmf}) to estimate the biased probability $\rho_{ij,kl}$, which is normalized counts $n_{ij,kl}$ up to a constant. We can feed these effective counts to the WHAM equations (Eq.~\ref{eq:C6}) to estimate $\rho_{ij,kl}$.\par
Although the first way is mathematically more rigorous, the second method is more convenient and cheaper to implement, and hence we did not use the first approach in this paper.

\section{Amount of simulation time used in each example}
\label{sec:simulation_table}

\begin{table}[h!]
\begin{ruledtabular}
\caption{Table of computational cost for original data of each example given in the main text.}
\label{tab:cost}
\begin{tabular}{c c c c c c}
System & Figure & Method & Cost/start & Starts & Windows \\
\hline
Harmonic & 1,2 & Quench & 10 + 2/$\gamma_\mathrm{quench}$ & 2000 & -  \\
Ala & 3a,4a,7 & 2dUS & 1 ns  & 1 & 400 \\
Ala & 3b & Quench & 1+4$\gamma^{-1}=$41 ps & $10^4$ & -  \\ 
Ala & 4b & Quench+2dUS & 3 ps & 300 & 400  \\ 
Ala & 5a,5c & 2dUS & 2 ns  & 1 & 400 \\
Ala & 5b & Quench+2dUS & 3 ps & 50 & 400  \\ 
Ala & 6a & 1dUS & 2 ns & 1 & 20  \\ 
Ala & 6b & Quench+1dUS & 3 ps & 666 & 20  \\ 
Ala & 7 & Quench+2dUS &  &  &   \\ 
 &  & $T_0=300$ & 3 ps & 666 & 400  \\ 
  &  & $T_0=75$ & 4 ps & 500 & 400  \\ 
  &  & $T_0=1200$ & 4 ps& 500 & 400  \\ 
Ala+H$_2$O & 8a & 2dUS & 2 ns & 1 & 400  \\ 
Ala+H$_2$O & 8b & S-Quench+2dUS & 3 ps & 333 & 400  \\ 
\end{tabular}
\end{ruledtabular}

\end{table}

\bibliographystyle{apsrev4-1}
\bibliography{quench}

\clearpage
\onecolumngrid
\section*{Supplementary Information}
\renewcommand{\thefigure}{S\arabic{figure}}
\setcounter{figure}{0}
\renewcommand{\thetable}{S\arabic{table}}
\setcounter{table}{0}
\renewcommand{\theequation}{S\arabic{equation}}
\setcounter{equation}{0}

\renewcommand{\thesection}{S\arabic{section}}
\setcounter{section}{0}

\section{Convergence of approximation in Eq.~\ref{eq:quenchpmf2}}

\begin{figure}[h]
    \centering
    \includegraphics{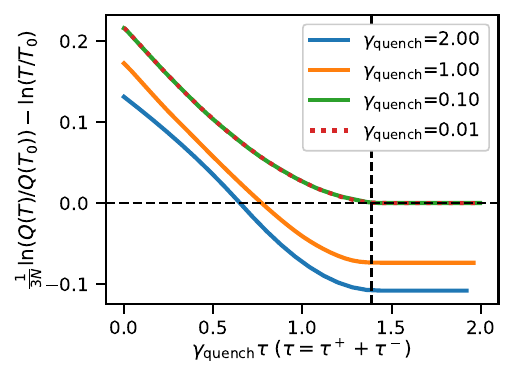}
    \caption{Error in the log of the ratio of partition functions at two different temperatures as a function of total quenching time for $\mathscr{N}=1000$ 3D harmonic oscillators. Quench is performed from 2000 initial samples obtained at $T_0=2.0$, estimating the partition function at $T=1$.  
    For small $\gamma_{quench}$, the partition function converges exactly to the theoretical value (horizontal dashed line shows zero error) at $\gamma_\mathrm{quench}\tau =2 \log(T/T_0)$ (vertical dashed line). When the quench is heavily damped, the ratio does not converge to the correct value.}
    \label{fig:Q_ratio_gt}
\end{figure}

\begin{figure}[h]
    \centering
    \includegraphics{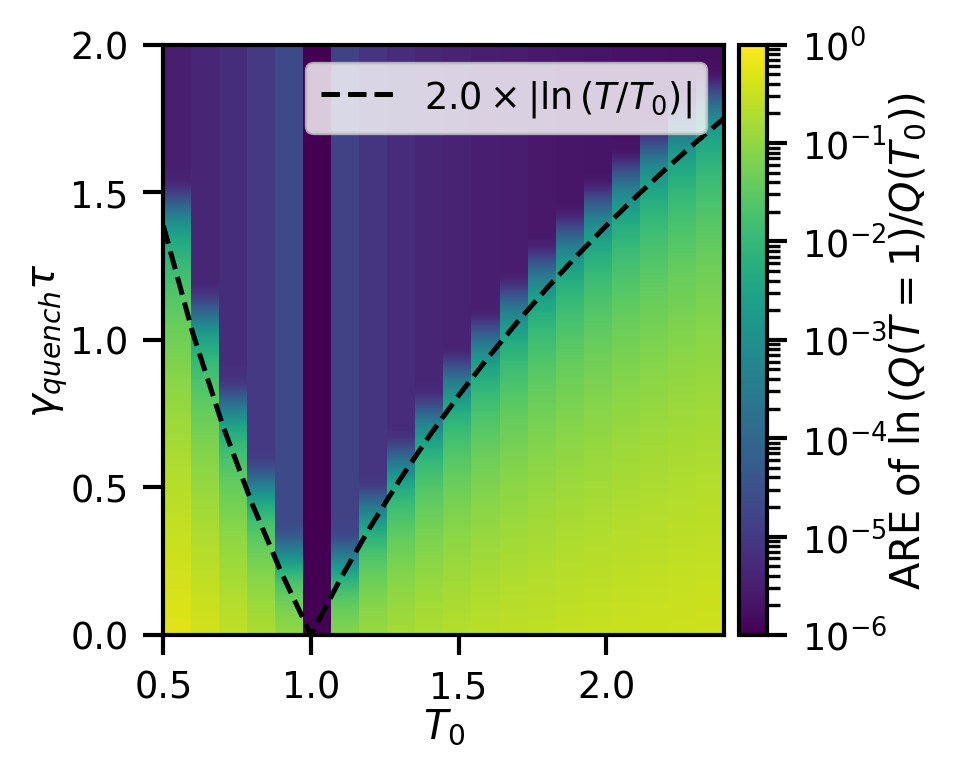}
    \caption{Absolute relative error of the log of the ratio of the partition function at $T=1$ with varying $T_0$, using $\gamma_{quench}=0.01$ ps$^{-1}$. 
    As $T_0$ goes away from 1, more sampling time $\gamma_\mathrm{quench}\tau$ is required.
    Total simulation time $\gamma_\mathrm{quench}\tau$ indicates quenching forward for $\gamma_\mathrm{quench}\tau/2+\ln{\left(T_0/T \right)}$ and backwards for  $-\gamma_\mathrm{quench}\tau/2+\ln{\left(T_0/T \right)}$.
    Dashed line shows the shape of $2 \ln{\left(\beta/\beta_0\right)}$, which is the time of convergence shown in Fig.~\ref{fig:Q_ratio_gt} . 
    }
    \label{fig:error_rt_gt}
\end{figure}

Fig.~\ref{fig:Q_ratio_gt} shows the convergence of the ``infinite time'' limit of our quench estimator is assessed for a system of harmonic springs in Fig.~\ref{fig:Q_ratio_gt} as described in the main text. We also assess the accuracy considering a range of starting temperatures and target temperatures, simulating for a range of times. 
To make the comparisons fair, we adjust the forward and backward time automatically by taking  $\gamma \tau^{+}=\gamma\tau^{+}_0+\ln{(T_0/T)}$ and $\gamma\tau^-=\gamma\tau^{-}_0+\ln{(T_0/T)}$ so that total simulation time is fixed and the overall temperature range is similar.
We show in Fig.~\ref{fig:error_rt_gt} that accurate results are obtained once total simulation time exceeds our heuristic value of $\gamma_\textrm{quench}(\tau^+-\tau^-)=\gamma_\textrm{quench}\tau=2|\ln(T_0/T)|$.

\section{Comparison of accuracy of the estimator in Eq.~\ref{eq:Q_ratio_quench} with the infinite time approximation and Cao and Vanden-Eijnden variants for harmonic springs}
\begin{figure}[h!]
    \centering
    \includegraphics{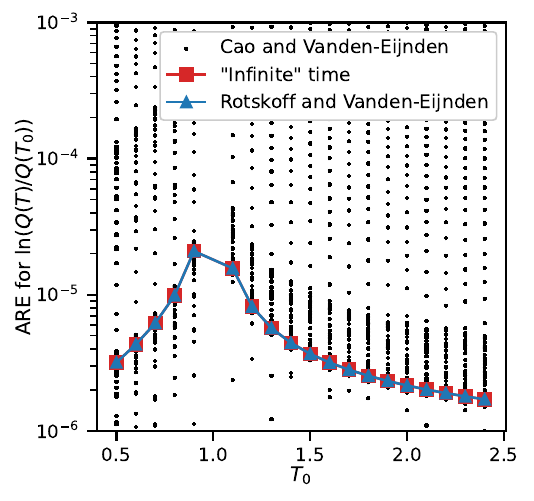}
    \caption{Absolute relative error of the log of the partition function ratio for $N=1000$ and $\gamma=0.01$ as in Fig.~\ref{fig:harmonic_vary_N_T} }
    \label{fig:compare_estimators_harmonic}
\end{figure}

In Fig.~\ref{fig:compare_estimators_harmonic} we compare the accuracy of estimating the ratio of partition functions for $N=1000$ harmonic springs with quench in two ways, (1) the infinite time limit using fixed total sampling time $\tau^+=2\gamma^{-1},\tau^-=-2\gamma^{-1}$ and (2) the finite time version of  Eq.~\ref{eq:general_expression_cao} from Ref.~\citenum{cao_learning_2022} for many different combinations of $\tau^+$ and $\tau^{-}$. 
The infinite limit gives equivalent accuracy, while the estimator from Ref.~\citenum{cao_learning_2022} can give higher accuracy results for specially chosen quench times, but is less accurate for typical choices, using total simulation time 4 $\gamma^{-1}$.

\section{Energy cutoffs for alanine quench}
\begin{figure}[h!]
    \centering
    \includegraphics{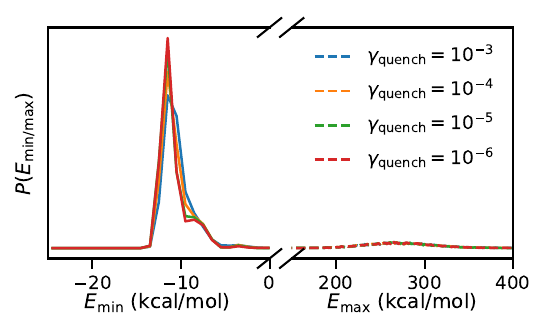}
    \caption{Energy histograms for fixed length forward and reverse quenches for alanine dipeptide. The minimum of the upper energies and the maximum of the lower energies were used as cutoffs for Fig.~\ref{fig:compare_ref_quench}. Here we started with $T_0=1200$K using $\gamma_\mathrm{quench}=1\times10^{-4}$ ps$^{-1}$ and $10^4$ starting points corresponding to $\approx 251$ ns of total simulation time. We chose $T_\mathrm{min}=200$K as a target lower quench temperature such that, $\gamma \tau^+=\ln(1200/200)=\ln(6)\approx 1.8$, and $ \gamma \tau^-=-2+\ln{(1200/300)}=-2+\ln(4)\approx -0.6$. }
    \label{fig:adp_energy_histogram}
\end{figure}

\section{Quench alone at different rates for alanine dipeptide}

\begin{figure}[h!]
    \centering
    \includegraphics[width=0.9\columnwidth]{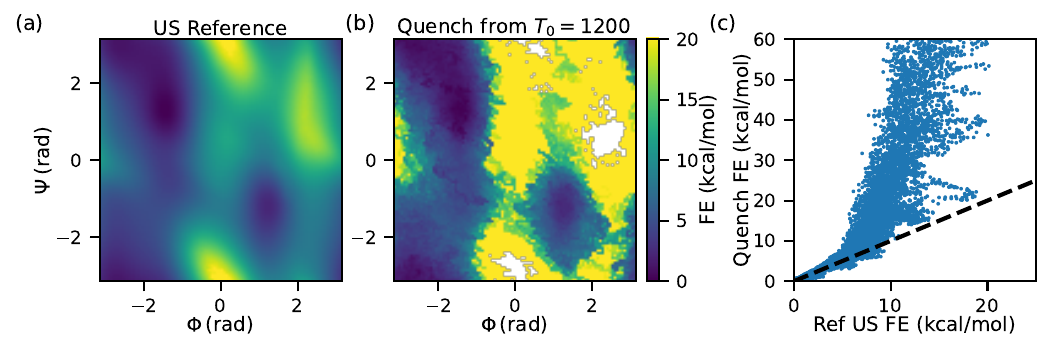}
    \caption{This shows the same procedure as in Fig.~\ref{fig:compare_ref_quench}, but with $\gamma_\mathrm{quench}=0.001$ ps$^{-1}$  (10 times faster). 
    Here, the total amount of sampling time in US is 400 ns and in quench is $\approx 50$ ns. Data used in computing Eq.~\ref{eq:quenchpmf} including generating restart points totals $\approx 33.5$ ns. }
    \label{fig:adp_fast_quench}
\end{figure}

\begin{figure}[h!]
    \centering
    \includegraphics[width=0.9\columnwidth]{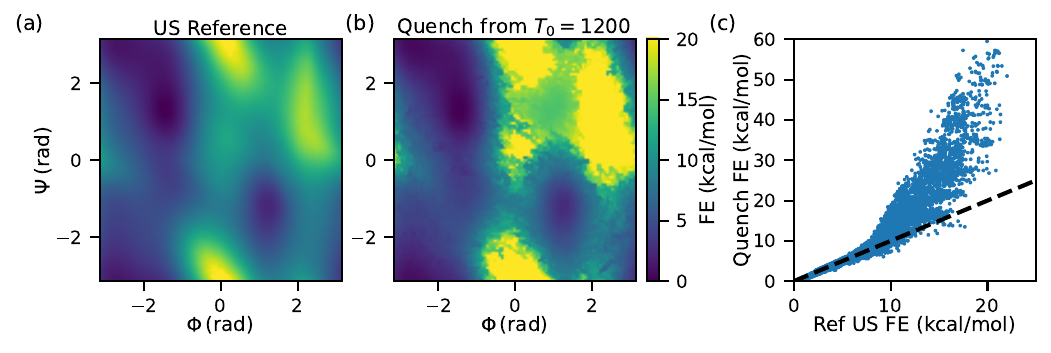}
    \caption{This shows the same procedure as in Fig.~\ref{fig:compare_ref_quench}, but with $\gamma_\mathrm{quench}=1\times10^{-5}$ ps$^{-1}$ (10 times slower). 
    Here, the total amount of sampling time in US is 400 ns and in quench is $\approx 4010$ ns. Data used in computing Eq.~\ref{eq:quenchpmf} including generating restart points totals $\approx 2529$ ns. }
    \label{fig:adp_slow_quench}
\end{figure}

\begin{figure}[h!]
    \centering    \includegraphics[width=0.9\columnwidth]{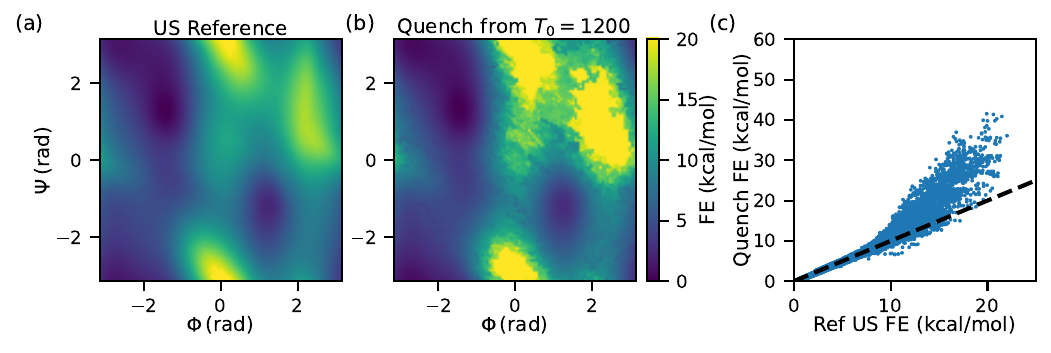}
    \caption{This shows the same procedure as in Fig.~\ref{fig:compare_ref_quench}, but with $\gamma_\mathrm{quench}=1\times10^{-6}$ ps$^{-1}$  (100 times slower). 
    Here, the total amount of sampling time in US is 400 ns and in quench is $\approx 40010$ ns. Data used in computing Eq.~\ref{eq:quenchpmf} including generating restart points totals $\approx 23997$ ns.} 
    \label{fig:adp_extraslow_quench}
\end{figure}

\clearpage
\section{Evaluation of FES computed by Quench using EMUS instead of WHAM for alanine dipeptide}
\label{sec:emus}

Here we show how EMUS\cite{thiede2016eigenvector} works in a normal umbrella sampling simulation and then derive a ``quench'' version of EMUS using Eq.~\ref{eq:quench_expression}.
In the EMUS algorithm, we first estimate three quantities by sample means:
\begin{equation}
    \left \{ \begin{aligned}
    \left\langle{\frac{\phi}{\sum_{k}{\psi_k}}} \right\rangle_i \equiv \langle \phi^* \rangle_i &\approx \frac{1}{N_i}\sum_{t=0}^{N_i-1}{\frac{\phi\left({{\bf X}_i(t)}\right)}{\sum_{k}{\psi_k\left({{\bf X}_i(t)}\right)}}}\\
    \left\langle{\frac{1}{\sum_{k}{\psi_k}}}\right\rangle_i \equiv \langle 1^* \rangle_i &\approx \frac{1}{N_i}\sum_{t=0}^{N_i-1}{\frac{1}{\sum_{k}{\psi_k\left({{\bf X}_i(t)}\right)}}}\\
    F_{ij} = \langle F_j \rangle_i &\approx \frac{1}{N_i}\sum_{t=0}^{N_i-1}{\frac{\psi_j\left({{\bf X}_i(t)}\right)}{\sum_{k}{\psi_k\left({{\bf X}_i(t)}\right)}}}
    \end{aligned} \right.,
\label{eq:E1}
\end{equation}
where $g$ is an observable of interest, $\psi_k\left({{\bf X}_i(t)}\right)=\frac{1}{2}\kappa\left({S({\bf X}_i(t))-s_k}\right)^2$ is the exact biased potential given sample data ${\bf X}_i(t)$ from the $i$th biased simulation as if it were sampled from the $k$th biased simulation. $\bf F$ is the overlap matrix and $F_{ij}$ is the element of $\bf F$ in $i$th row and $j$th column. With these definitions, we can compute expectations of $\phi$ by the following expression:

\begin{equation}
    \langle{\phi}\rangle = \frac{\sum_{i=1}^{L}{z_i\langle{\phi^*}\rangle_i}}{\sum_{i=1}^{L}{z_i\langle{1^*}\rangle_i}},
\label{eq:E2}
\end{equation}
where $z_i$ solves

\begin{equation}
    z_j = \sum_{i=1}^{L}{z_i F_{ij}}.
\label{eq:E3}
\end{equation}
and $z$ is uniquely determined since

\begin{equation}
    \sum_{i=1}^{L}{z_i}=1.
\label{eq:E4}
\end{equation}

In the case of ``quench'' and umbrella sampling simulation, we use Eq.~\ref{eq:quench_expression} as a bridge to connect an equilibrium average and the data sampled out of equilibrium.
We first estimate three quantities as follows
\begin{equation}
    \left \{ \begin{aligned}
    \langle \phi^* \rangle_i &\approx \frac{1}{N}\sum_{k=1}^{N}{\frac{\int{\frac{\phi({\bf X}_k(t))}{\sum_{l=1}^{N_w}{\psi_l({\bf X}_k(t))}}e^{-\beta_0\mathscr{H}_i({\bf X}_k(t))-d\gamma t}\,dt}}{\int{e^{-\beta_0\mathscr{H}_i({\bf X}_k(t))-d\gamma t}\,dt}}}\\
    \langle 1^* \rangle_i &\approx \frac{1}{N}\sum_{k=1}^{N}{\frac{\int{\frac{1}{\sum_{l=1}^{N_w}{\psi_l({\bf X}_k(t))}}e^{-\beta_0\mathscr{H}_i({\bf X}_k(t))-d\gamma t}\,dt}}{\int{e^{-\beta_0\mathscr{H}_i({\bf X}_k(t))-d\gamma t}\,dt}}}\\
    \langle F_j \rangle_i &\approx \frac{1}{N}\sum_{k=1}^{N}{\frac{\int{\frac{\psi_j({\bf X}_k(t))}{\sum_{l=1}^{N_w}{\psi_l({\bf X}_k(t))}}e^{-\beta_0\mathscr{H}_i({\bf X}_k(t))-d\gamma t}\,dt}}{\int{e^{-\beta_0\mathscr{H}_i({\bf X}_k(t))-d\gamma t}\,dt}}}
    \end{aligned} \right.,
\label{eq:E5}
\end{equation}
Here, the integrals are computed from $\tau_k^-$ to $\tau_k^+$ as in the main text.

After computing these quantities, we find the left eigenvector $z_i$ that satisfies Eq.~\ref{eq:E3} and use Eq.~\ref{eq:E2} to estimate an arbitrary observable from our quench data. 
An FES is obtained if we choose $\phi=\delta(S({\bf X})-s)$.
Note that a numerical error is introduced since we have used Eq.~\ref{eq:quench_expression} as a bridge, although no numerical error is introduced when making grids as in the case of WHAM. 
Below, we demonstrate that EMUS does work for the quench results, as compared to an equilibrium EMUS calculation and as compared to WHAM. 
However, EMUS is much more computationally expensive so we have used WHAM to reweight data from ``quench'' and umbrella sampling simulations in practice.

\clearpage 
\begin{figure}[h!]
    \centering
    \includegraphics{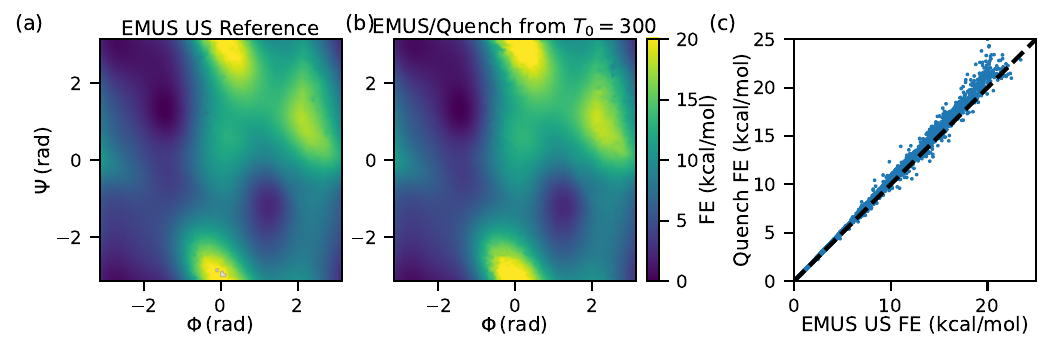}
    \caption{A comparison of FES between EMUS and ``quench'' version of EMUS. (a) FES computed from US at $T=300\textrm{K}$ with 800 ns total sampling time. (b) FES computed from ``quench'' version of EMUS derived in Section \ref{sec:emus} with $\gamma_{\textrm{quench}}=0.001$ ps$^{-1}$ and 863.1 ns total simulation time.}
    \label{fig:emus_si_1}
\end{figure}
\begin{figure}[h!]
    \centering
    \includegraphics{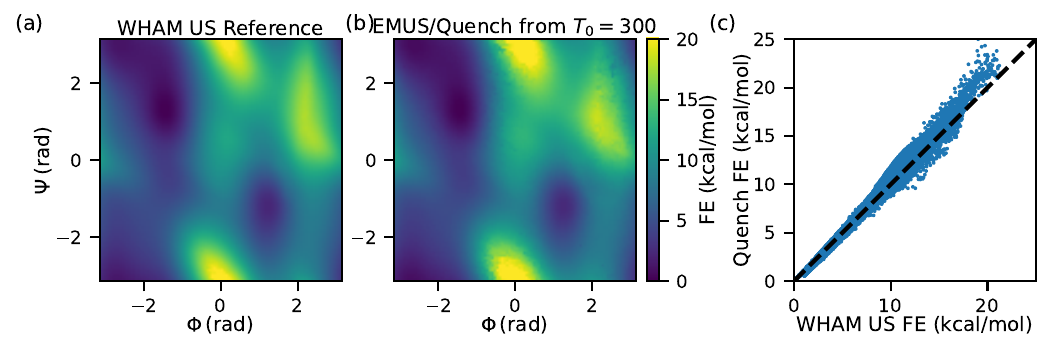}
    \caption{Same as Fig.~\ref{fig:emus_si_1}, but comparing Quench+US/EMUS to the FES computed by WHAM as in the main text.}
    \label{fig:emus_si_2}
\end{figure}

\clearpage
\section{Comparison of FES computed by US+WHAM with EMUS and Metadynamics for alanine dipeptide}

\begin{figure}[h!]
    \centering
    \includegraphics{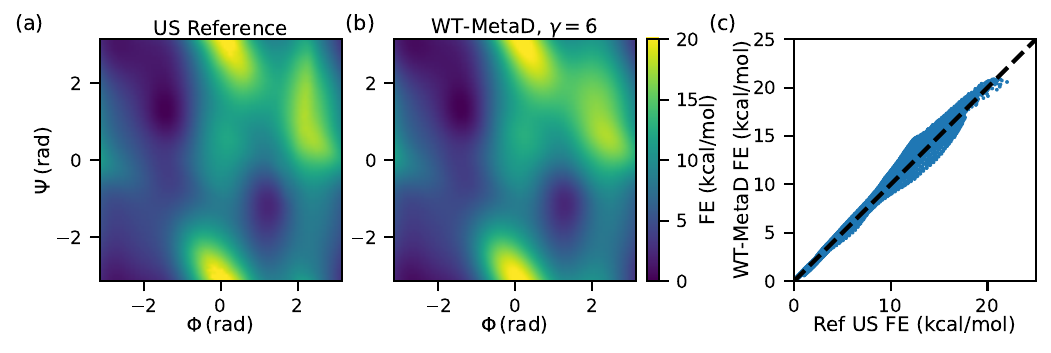}
    \caption{Sames as  Fig.~\ref{fig:compare_ref_quench}, but panel (b) is computed with WT-MetaD with a bias factor of 6, as described in the main text.}
    \label{fig:metad_bf6}
\end{figure}

\begin{figure}[h!]
    \centering
    \includegraphics{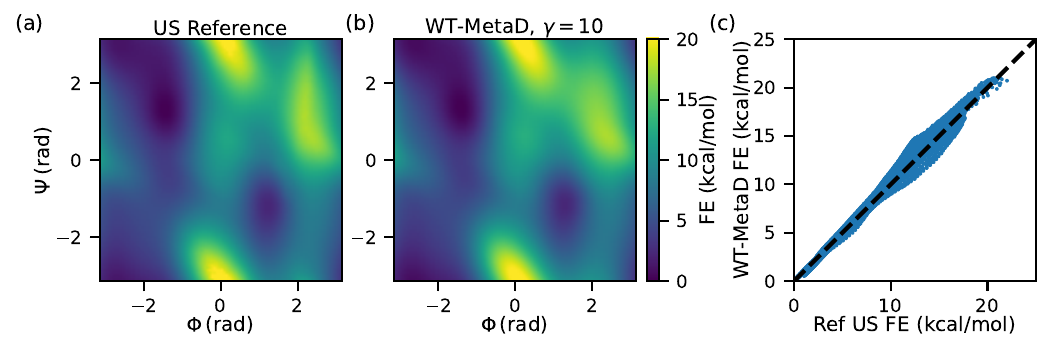}
    \caption{Sames as  Fig.~\ref{fig:compare_ref_quench}, but panel (b) is computed with WT-MetaD with a bias factor of 10, as described in the main text.}
    \label{fig:metad_bf10}
\end{figure}

\begin{figure}[h!]
    \centering
    \includegraphics{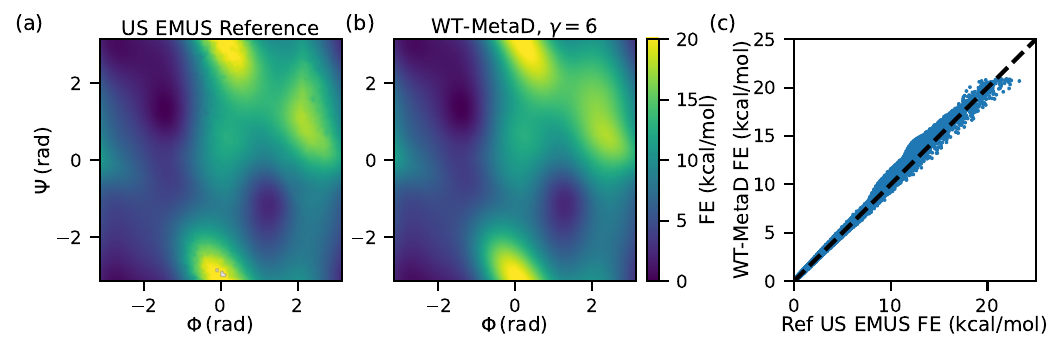}
    \caption{Sames as  Fig.~\ref{fig:metad_bf6}, but panel (a) is computed EMUS rather than WHAM.}
    \label{fig:emus_metad_bf6}
\end{figure}

\end{document}